\documentclass[superscriptaddress,twocolumn,bibnotes,
amsmath,amssymb,aps,pra,floatfix]{revtex4-2} 

\usepackage{graphicx}
\usepackage[caption=false]{subfig}
\usepackage{pgfplots}
\usepackage{physics}
\DeclareFontFamily{OT1}{pzc}{}
\DeclareFontShape{OT1}{pzc}{m}{it}{<-> s * [1.10] pzcmi7t}{}
\DeclareMathAlphabet{\mathpzc}{OT1}{pzc}{m}{it}
\usepackage{dcolumn}
\usepackage{bm}
\usepackage{hyperref}
\usepackage{xcolor}
\usepackage{mathtools}
\usepackage{tikz}
\usepackage{float}
\usetikzlibrary{patterns}
\usepackage{cases}
\usepackage{soul}
\newcommand\underrel[3][]{\mathrel{\mathop{#3}\limits_{%
      \ifx c#1\relax\mathclap{#2}\else#2\fi}}}

\usepackage{natbib}
\usepackage{braket}

\definecolor{darkgreen}{rgb}{0.1020,0.6941,0.2667}

\begin{document}


\title{Resonance-facilitated three-channel p-wave scattering}
\author{Denise J.\ M.\ Ahmed-Braun}
\affiliation{Department of Physics, Eindhoven University of Technology, The Netherlands}
\author{Paul S.\ Julienne}
\affiliation{Joint Quantum Institute, NIST and the University of Maryland, U.S.A.}
\author{Servaas J.\ J.\ M.\ F.\ Kokkelmans}
\affiliation{Department of Physics, Eindhoven University of Technology, The Netherlands}

\date{\today}

\begin{abstract}
Feshbach resonances of arbitrary width are typically described in terms of two-channel models. 
Within these models, one usually considers a single dressed resonance, with the option to extend the analysis by including resonant open-channel features that can drastically change the observed threshold effects. For the strong $^{40}\mathrm{K}$ p-wave resonance studied in Ref. \cite{ahmed2021}, the interplay between an open-channel shape resonance and the Feshbach resonance could explain the unexpected nonlinear variation of the binding energy with magnetic field.  However, the presented two-channel treatment relies on the introduction of two independent fitting parameters, whereas the typical Breit-Wigner expression would only account for one. This results in an effective magnetic moment that acquires a nonphysical value, which is an indication of a major shortcoming of the two-channel model treatment. In this study, we observe how the presence of a closed-channel shape resonance explains the physical mechanism behind the observations and demonstrates the need of a three-channel treatment. We introduce our novel model as \textit{resonance facilitated}, where all coupling is mediated by the Feshbach state, while there is no direct coupling between the additional channel and the open channel. Notably, the resonance-facilitated structure greatly reduces the complexity of the full three-channel model. The typical Breit-Wigner form of the two-channel Feshbach formalism is retained and the full effect of the added channel can be captured by a single resonance dressing factor, which describes how the free propagation in the Feshbach state is dressed by the added channel.  
\end{abstract}

\pacs{Valid PACS appear here}
\maketitle

\section{\label{sec:level1}Introduction}
Feshbach resonances have given experimentalists unprecedented control over two-body interactions in quantum degenerate fluids, greatly adding to the versatility of quantum gases.
The tunability of Feshbach resonances follows from the magnetic moment difference $\delta\mu$ between hyperfine channels. Whereas the multichannel nature of these resonances is essential, they are complex in nature.
The easiest approach to retain the magnetic-field dependence is to use a two-channel model. Especially for pairs formed in s-wave channels, these models have been successful in describing both resonance width-dependent as well as universal behavior \cite{art:chin}. 
\par
Following the experimental observation of p-wave resonances, \cite{art:regal_cond,gunter2005,gaebler2007,zhang2004,schunck2005,fuchs2008,nakasuji2013,waseem2017,SalomonThreeChannel,jochim2003,zwierlein2003,bourdel2004,kinast2004,regal2004}, recent studies have revealed the existence of p-wave universal behavior \cite{Luciuk2016,yoshida2015,yu2015,he2016,peng2016}. These systems are considerably different from their s-wave counterparts. Despite the presence of strong three-body losses, they exhibit interesting features. 
P-wave interactions allow for the existence of multiple superfluid phases related to different projections of the angular momentum of Cooper pairs, some of which are similar to the phases of superfluid He-3 \cite{vollhardt2013}, and exhibit a phase transition if tuned from BCS to BEC rather than a smooth crossover \cite{volovik1992}.  
In addition, the prospects of duality between strongly interacting odd waves and weakly interacting even waves in one-dimensional systems with suppressed three-body losses \cite{cheon1999,girardeau2005,girardeau2006,sekino2018} and the topological phase transitions in two-dimensional systems \cite{read2000,tewari2007} and engineered states \cite{jiang2016,yang2016,hu2016} explain the interest in understanding the details of p-wave interactions. \par
\begin{figure}
\begin{center}
\includegraphics[width=0.7\columnwidth]{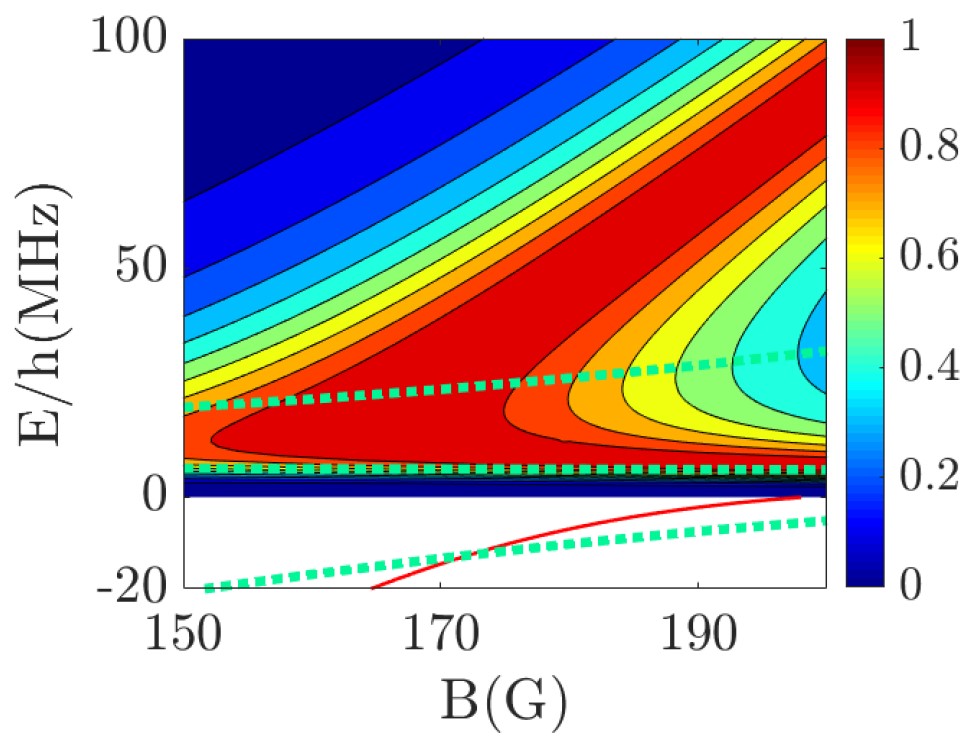}
\caption{\label{fig:diffmu_2channel} \textbf{Binding energy and inelastic loss in the $\ket{bb}$ channel}
The below threshold binding energy $E_b$ (full red line) and above threshold inelastic loss (contour plot) of the $\ket{bb}$ state as a function of magnetic field $B$ are extracted from CC calculations. The green dashed line follows from the artificial two-channel model mapped back onto the original Breit-Wigner form. Whereas it captures the bending of the dimer energy near-threshold consistent with CC data, it fails to match the data quantitatively due to the poorly fitted differential magnetic moment $\delta \mu$.}
\end{center}
\end{figure}

 
The explicit two-channel treatment of p-wave resonances has been the topic of several theoretical studies \cite{SalomonThreeChannel,gubbels2007,levinsen2008}.
Similar to their s-wave counterparts, these two-channel models allow for the inclusion of resonant open-channel features and can correctly capture the interplay between open-channel and Feshbach resonances \cite{marcelis2004,ahmed2021}. However, since the Feshbach channel in these models can comprise multiple hyperfine channels, resonant features in these hyperfine channels cannot be treated explicitly in this formalism. 
The breakdown of the two-channel model can be observed in the $^{40}\mathrm{K}$ Feshbach resonance studied in Ref.\cite{ahmed2021}. 
Here, the Feshbach part of the two-channel scattering matrix, or S-matrix, element $S_{\mathrm{FB}}$ had to be redefined in order to match coupled-channels (CC) data. The typical Breit-Wigner form of the S-matrix element \cite{mott1965theory,taylor2006} was replaced by a function where the resonance width and shift could be fitted to the CC data independently, such that
\begin{align}
\label{eq:2artificial}
S_{\mathrm{FB}} &= 1-\frac{i \Gamma(E)}{E-\delta\mu (B-B_n)-\Delta_{\mathrm{res}}(E)+\frac{i}{2} \Gamma(E)} \notag \\ 
&\rightarrow 1- \frac{i g k^3}{E-c+ \frac{i}{2} g k^3}
\end{align}
Whereas the two-channel model can capture the correct low-energy scaling, the differential magnetic moment $\delta\mu \approx 0.5 \mathrm{G/MHz}$ that can be extracted from the fit by mapping the artificial model back onto the Breit-Wigner form, results in a poor quantitative match of the realistic two-channel model with the CC data as shown in Fig. 3(c) of Ref.~\cite{ahmed2021} and reproduced here in Fig~\ref{fig:diffmu_2channel}. \par 
In this paper we relate this discrepancy to the presence of a near-threshold shape resonance in one of the hyperfine channels other than the entrance channel. The need to treat this channel explicitly and hence upgrade to a three-channel model to correctly capture the physics is not specific to Feshbach resonances. 
Other fields of physics, including two-color photo-association experiments, Stimulated Raman adiabatic passage (STIRAP) \cite{ates2007,heeg2012} as well as electromagnetic induced transparency (EIT) \cite{ates2011,petrosyan2011,peyronel2012,sevinccli2011} similarly reveal new physics that cannot be explained by reduced two-channel models. \par 
The inclusion of extra channels to the scattering problem rapidly increases the complexity of the analysis and risks obscuring the intuitive understanding of the physics that two-channel models offer.
As motivated in Sec.~\ref{sec:Inelastic loss}, the magnetic-field dependence of the inelastic loss in the $\ket{bb}$ channel presented in Fig.~\ref{fig:diffmu_2channel} leads us to expect that the system is well-described in terms of a model where all interactions are facilitated by the Feshbach resonance. 
This resonance-facilitated structure reduces the complexity significantly. It allows the effect of the third channel to be fully captured by a single dressing factor, later defined as $\mathrm{D}$. Physically, this factor describes how the free propagation in the Feshbach state is dressed by the third channel. As such, we retain the overall Breit-Wigner form of the two-channel S-matrix, allowing for a relatively straightforward physical interpretation of the results as compared to full CC output. We find that the resonance-facilitated three-channel model with shape resonances in the entrance channel and the added third channel contains the correct low-energy physics and, as outlined in Sec.~\ref{subsec:CC structure}, allows for the direct implementation of values for $\delta\mu$ consistent with CC data.  \par 
This paper is outlined as follows. In Sec.~\ref{sec:CC calculations} we study the CC structure of the relevant $^{40}\mathrm{K}$ system and use full CC data to motivate the reduction to the resonance-facilitated model. Next, in Sec.~\ref{sec:Factorisation of the S-matrix} we analyse how the presence of open-channel resonances can generally affect the threshold behavior of a ramping p-wave Feshbach state. We then proceed with the derivation of the resonance facilitated three-channel model in Sec.~\ref{sec:Resonance facilitated scattering}, and discuss how we use a Gamow expansion to account for the shape resonances. In Sec.~\ref{sec:ResonanceWidth3CH}, we present the field dependence of the resonance scattering parameters. These parameters will then be used to interpret the results in Sec.~\ref{sec:Results}, compute the resonance width and to form the conclusions in Sec.~\ref{sec:conclusion}.
 \par 
  \section{ \label{sec:CC calculations}Coupled-channels calculations}
We study fermionic $^{40}\mathrm{K}$, which has a nuclear spin of four and ground state $^2\mathrm{S}_{\frac{1}{2}}$. Hence, the single particle hyperfine ground state manifold contains the total spin states $f = 9/2$ and $f=7/2$, with respectively ten and eight spin components with projections $m_{f}$. 
We label these states as $\ket{a},\ket{b},\ket{c}, ...$ in order of increasing energy. As $^{40}\mathrm{K}$ has an inverted hyperfine structure, the entrance channel of interest in this paper with two atoms in the $\ket{f,m_f} = \ket{9/2,-7/2}$ state at zero $B$-field corresponds to the $\ket{bb}$ channel. Apart from the Feshbach state, which is a magnetic-field dependent combination of hyperfine channels, we explicitly include the $\ket{ac}$ channel in the three-channel model. 
\subsection{\label{subsec:CC structure}{Coupled-channels structure}}
\begin{figure}[H]
\begin{center}
\includegraphics[width=\columnwidth]{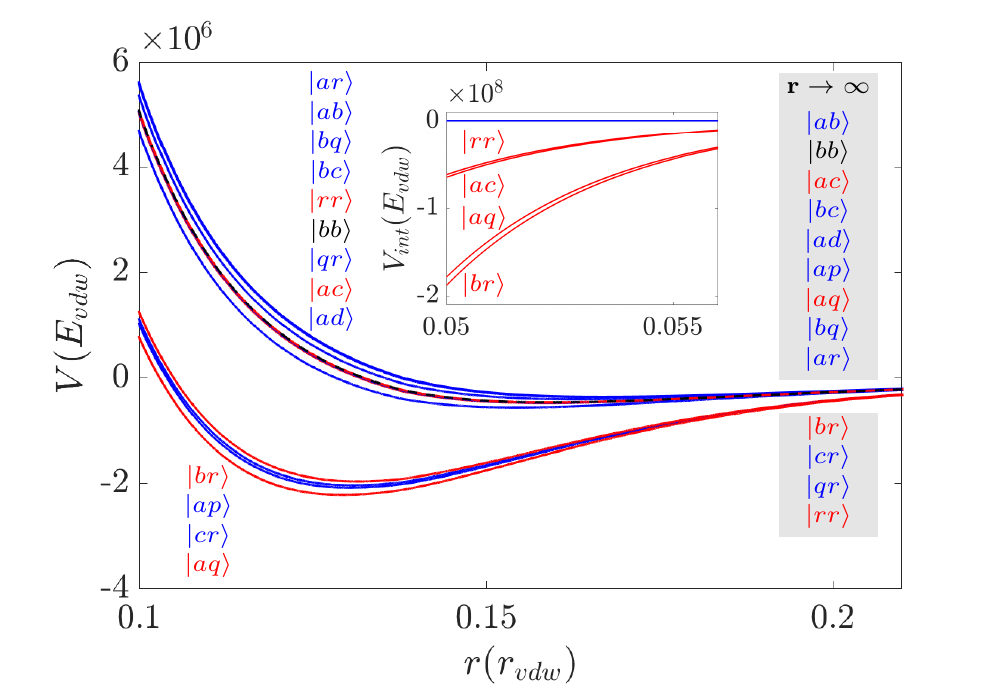}
\caption{\textbf{Channel and interaction- potentials.} Channel potentials $V_{\mathrm{ch},\mathrm{ch}}$ in the $M_{\mathrm{tot}}=-7$ subspace expressed in units of the van der Waals (vdW) energy $E_{vdW}/h = 23.375 \mathrm{MHz}$ as a function of the inperparticle distance in terms of the vdW length $r_{vdW} =  65.0223 a_0$, with Bohr radius $a_0$. 
Channels are either coupled to the $\ket{bb}$ entrance channel (black line) through spin-exchange interactions (red lines) or dipole-dipole interactions (blue lines). All channels are listed on the right side of the figure in order of decreasing asymptotic energy and on the left side of the figure in order of potential energy in the plotted regime. Here, the top group of potentials is predominantly triplet, whereas the bottom group of potentials is predominantly singlet. Using the same color coding, we indicate the interaction potentials to the $\ket{bb}$ channel $V_{\mathrm{bb},\mathrm{ch}}$ in the inset. Only channels directly coupled to the $\ket{bb}$ state through spin exchange interactions have a distinguishable value on this scale. \label{fig:PotentialsMtotm7}} 
\end{center}
\end{figure}

To account for the relative angular momentum between two interacting atoms, we extend the two-particle hyperfine basis by including the partial wave quantum numbers $L$ and $M_{L}$.
Since we are interested in the collision between two atoms which are both in the $\ket{b}$ state, the antisymmetry requirement for fermions implies that these atoms can only collide with odd $L$ values. In the low-energy limit, this results in the dominance of p-wave interactions with $L=1$ and $M_{L} = -1,0$ or $1$. 
At small inter-particle separations, the $\ket{b}$ atoms experience direct spin-exchange interactions that couple hyperfine channels with conserved $m_{f_1}+m_{f_2}$. However, the anisotropy of the p-wave interaction additionally results in a non-zero dipole-dipole interaction that couples additional channels with conserved $M_{\mathrm{tot}} = m_{f_1}+m_{f_2}+M_{L}$.
This means that the collision channels with $M_{\mathrm{tot}} = -8,-7$ and $-6$ couple $8,13$ and $20$ channels respectively.
Figure \ref{fig:PotentialsMtotm7} shows the channel potentials that are coupled for interactions with $M_{\mathrm{tot}} = -7$. Whereas the inset reveals that the interaction potential $V_{\mathrm{int}}$ is much larger for direct spin-exchange interactions, the dipole-dipole interaction can generally not be neglected and results in the experimentally observable splitting of the Feshbach resonance for different values of $\abs{M_L}$ \cite{Ticknor:2004}. 

For all values of $M_{\mathrm{tot}}$, there are five channels that couple through direct spin-exchange interactions. As can be seen in Fig.~\ref{fig:PotentialsMtotm7}, three of these channels, $\ket{bb}$, $\ket{rr}$ and $\ket{ac}$, are predominantly triplet in character, whereas the channels $\ket{aq}$ and $\ket{br}$ are mostly singlet. 
The inset reveals that the coupling between the singlet channels and the $\ket{bb}$ state is about three times larger than the coupling between the triplet channels and the $\ket{bb}$ state. This is consistent with the spin-exchange interaction following from the energy difference between the singlet and triplet potentials, such that it should be larger for states with predominantly different spin symmetry. \par
Similarly to the $\ket{bb}$ state, the $\ket{ac}$ channel has a near-threshold shape resonance. It is the interplay between this additional shape resonance and the Feshbach state which we aim to capture by going from a two- to a three-channel model.
\subsection{\label{sec:Inelastic loss}The differential magnetic moment and inelastic losses}
Around resonance ($B \approx 198.3/8$ G) the channels $\ket{bb}$ and $\ket{ac}$ are separated by an asymptotic energy difference of $E_{\mathrm{th}}/h \approx 2.4$ MHz. Defining the energy $E$ of the incoming state with respect to the $\ket{bb}$ threshold, this means that $\ket{ac}$ is energetically closed for $E < E_{\mathrm{th}}$ and open for $E \geq E_{\mathrm{th}}$. 
Figure~\ref{fig:diffmu_2channel} shows the rapid increase in the $\ket{bb}$ channel inelastic loss once the $\ket{ac}$ channel opens.
The shape resonances in the $\ket{bb}$ and $\ket{ac}$ are magnetic-field independent and the energy difference $E_{\mathrm{th}}$ scales with the magnetic moment difference between $\ket{bb}$ and $\ket{ac}$, which is small \footnote{This is expected for two channels that are both predominantly triplet and is clearly Visible in Fig.3b of Ref. \cite{ahmed2021}}. This means that loss caused by direct coupling between the two shape resonances should be largely magnetic-field independent. 
However, Fig.~\ref{fig:diffmu_2channel} instead shows that the observed loss feature is strongly magnetic-field dependent. 
This implies the importance of the ramping Feshbach state and motivates the usage of the resonance facilitated three-channel model where $\ket{bb}$ and $\ket{ac}$ are solely coupled to the Feshbach state. \par
\begin{figure}[H]
\centering
\includegraphics[width=\columnwidth]{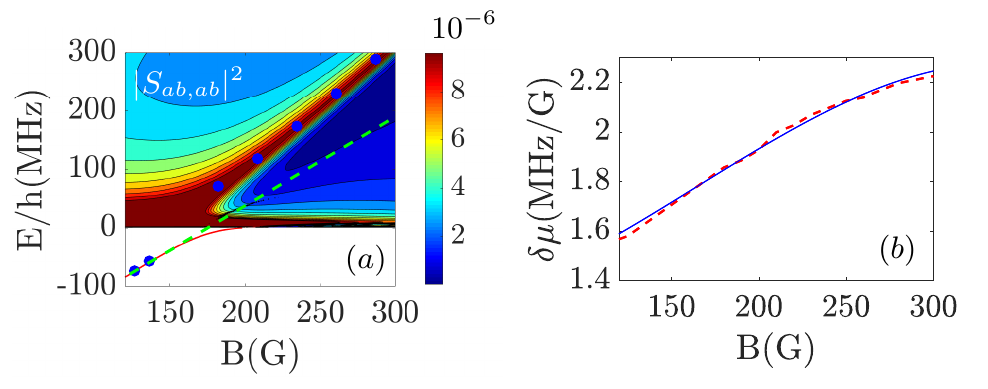}
\caption{\label{fig:Contour_diffmu180to340}{\bf Computing the differential magnetic moment} (a) The below-threshold binding energy (red line) and above threshold value of $\abs{S_{ab,ab}}^2$. The $\ket{ab}$ state is weakly interacting with $\ket{bb}$ and is therefore an excellent probe of $\delta\mu$. The blue circles indicate the localization of the binding energy/loss maxima. The dashed green line indicates how the slope $dE/dB$ is significantly different below/above threshold. (b) The dashed red line shows the differential magnetic moment computed using the binding energy values and loss maxima as a function of B-field. The blue line represents a quadratic fit of the data points. The values of this fit are used to define $\delta\mu(B)$.} 
\end{figure}

The magnetic-field dependence of the Feshbach channel is quantified by $\delta\mu$. Away from resonance, where the dressing of the (quasi)-bound state by the $\ket{bb}$ and $\ket{ac}$ channels is negligible, $\delta\mu$ can be calculated from the slope of the (quasi)-bound state energy versus magnetic-field. 
Connecting the below- and above- threshold regions as visualized in Fig.~\ref{fig:Contour_diffmu180to340}(a) through interpolation, we find the magnetic-field dependent $\delta\mu$ as presented in Fig.~\ref{fig:Contour_diffmu180to340}(b) \footnote{The B-field dependence of $\delta\mu$ is also visible in Fig.3(b) of Ref.~\cite{ahmed2021}, where the value of $\mu(B)$ for various hyperfine channels is presented. In the case of constant $\delta\mu$, the lines in Fig. 3b should remain parallel. This is however clearly not the case in the considered B-field regime.}.
The interpolated values for $\delta\mu$ will be used in the comparison of the resonance facilitated three-channel model to the CC data. 
Hence, contrary to the method followed in Ref. \cite{ahmed2021}, $\delta\mu$ is not a free parameter in our analysis and has a physically realistic value. 

 \section{\label{sec:Factorisation of the S-matrix}Poles and resonances of the S-matrix}
Considering the effect of the $\ket{bb}$ shape resonance, the Feshbach coupling and the dipole-dipole interaction separately, we express the total S-matrix as 
\begin{align}
S_{\mathrm{tot}} = S_{\mathrm{P}}S_{\mathrm{FB}}S_{\mathrm{dip}},
\end{align}
 where $S_{\mathrm{P}}$ represents the direct scattering part and where $S_{\mathrm{dip}}$ represents the dipole-dipole contribution. Using the Born approximation presented in Sec. IV C of Ref.~\cite{ahmed2021} to factor out the dipole-dipole contribution, the remaining S-matrix $S = S_{\mathrm{tot}} S_{\mathrm{dip}}^{-1}$ should follow the typical effective range approximation (ERA) for p-wave interactions as stated in Eq.~\eqref{eq:ERApwave} and will be used in the remainder of this paper. In the following two subsections, we proceed with the separate analysis of the remaining contributions $S_{\mathrm{P}}$ and $S_{\mathrm{FB}}$. 
\subsection{\label{subsec:Single-channel}Single channel}
\subsubsection{\label{subsec:Factorizing the S-matrix}Factorizing the S-matrix}
Solving the radial Sch\"odinger equation one can generally define $\psi_{\ell}(k,r)$ as the physical solution which satisfies two boundary conditions and which represents the radial part of the 
wave function of the normalized state $\ket{E,\ell,m}$. 
Whereas physically correct, the two boundary conditions imply that the normalization of the wave function is a function of the potential at all points. For physically realistic potentials this is often complex and one has to rely on numerical methods to compute the radial wave function. \par
Hence, one can alternatively define a different class of solutions which satisfy single point boundary conditions. Two of such solutions are the regularized wave function $\phi_{\ell}(k,r)$ and the Jost solutions $f_{\ell}(k,r)^{\pm}$, respectively defined as \cite{kukulin2013}
\begin{align}
&\underset{r\rightarrow 0}{\text{lim}} (2\ell+1)!! r^{-\ell-1}\phi_{\ell}(k,r) = 1 \\
&\underset{r\rightarrow \infty}{\text{lim}} e^{\pm ikr} f_{\ell}^{\pm}(k,r) = i^{\pm l}
\end{align}
The previous two expressions indicate that $\phi_{\ell}(k,r)$ is defined to be regular at the origin whereas the Jost solutions $f_{\ell}(k,r)^{\pm}$ are set to define purely incoming/outgoing waves, but are non-regular at the origin. \par
As both classes of solutions solve the Schr\"odinger equation, they can be matched by considering their asymptotic limits, finding that $\psi_{\ell}(k,r) = \phi_{\ell}(k,r)/\mathcal{F}_{\ell}(k)$ and more notably \cite{taylor2006}
\begin{align}
\label{eq:StoJost}
S_{\ell}(k) = \frac{\mathcal{F}_{\ell}(-k)}{\mathcal{F}_{\ell}(k)},
\end{align}
where we have introduced the Jost function
\begin{align}
\mathcal{F}_{\ell}(k) = \underset{r\rightarrow 0}{\text{lim}} \frac{(-kr)^{\ell}}{(2\ell-1)!!} f_{\ell}^{\pm}(k,r).
\end{align}
Equation \eqref{eq:StoJost} is particularly useful as the single point boundary condition which is used to define the Jost functions ensures that these functions can be solved iteratively for any potential $V(r)$ and allows for a Hadamard expansion \cite{regge1958}
\begin{align}
\label{eq:JostExpansion}
f_{\ell}(k,r) = f_{\ell}(0,r) e^{i k r_\mathrm{c}} \prod_{\mathrm{n}} \left(1-\frac{k_{\mathrm{n}}}{k}\right),
\end{align}
with momentum-space poles $k_{\mathrm{n}}$ and undetermined constants $f_{\ell}(0,r)$ and $r_{\mathrm{c}}$ which are set by non-resonant background scattering processes. 
Substituting Eq.~\eqref{eq:JostExpansion} into Eq.~\eqref{eq:StoJost}, we find the Ning-Hu representation of the S-matrix \cite{Ning:1948NingHuRepresentation}
\begin{align}
\label{eq:Sexpansion}
S_{\ell}(k) = e^{-2ik r_{c}} \prod_{\mathrm{n}} \left(\frac{k_{\mathrm{n}}+k}{k_{\mathrm{n}}-k}\right).
\end{align}
Equation \ref{eq:Sexpansion} enables us to express the full S-matrix in terms of its resonant poles. 
These resonant poles can be divided into four categories depending on their position in the complex momentum plane \cite{goosen2011}. \par
First of all, the poles located on the positive imaginary axis ($k_n = i\beta$, with $\beta > 0$) correspond to bound states. The wave function of these states decays exponentially in the asymptotic regime, such that $\psi_{\ell}(k_{\mathrm{n}},r) \underset{r\rightarrow \infty}{\sim}e^{-\beta r}$. Of all the four classes of poles, only these poles correspond to physical states.
The second class of poles are located the negative imaginary axis  ($k_n = -i\beta$) and are referred to as virtual, or anti-bound states. Contrary to the bound states, the asymptotic part of the wave function of these virtual states increases exponentially, such that $\psi_{\ell}(k_{\mathrm{n}},r) \underset{r\rightarrow \infty}{\sim}e^{\beta r}$. 
The third and fourth classes of poles are related as, according to the Schwarz reflection principle \cite{kukulin2013}, they always occur as twin poles located at $k =\pm \alpha - i\beta$ with $\alpha > 0 $. 
The resonance states ($k =\alpha - i\beta$) represent outgoing waves whose amplitude increase exponentially, such that $\psi_{\ell}(k_n,r)  \underset{r\rightarrow \infty}{\sim} e^{i \alpha r}e^{\beta r}$ and are hence also referred to as decaying states. 
The anti-resonance states on the other hand ($k =-\alpha - i\beta$) represent incoming waves with an asymptotic wave function of the form $\psi_{\ell}(k_n,r)  \underset{r\rightarrow \infty}{\sim} e^{-i \alpha r}e^{\beta r}$ and are hence also referred to as capturing states. \par 
\begin{figure}[h!]
\begin{center}
\includegraphics[width=0.9\columnwidth]{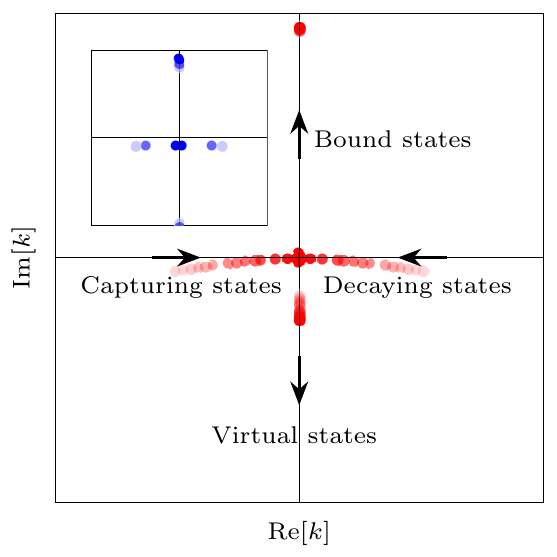}
\caption{\label{fig:PolesPicture} \textbf{Poles of the S-matrix} of the p-wave square-well in the complex momentum plane. Increasing color intensity corresponds to increasing depth of the potential well. The arrows indicate the direction in which the poles move. The inset shows the poles of the S-matrix of the s-wave square-well.}  
\end{center}
\end{figure}
Whereas all four classes of poles can occur for all partial wave states, the presence of a centrifugal barrier impacts the pole locations. As first studied for square-wells in Ref. \cite{nussenzveig1959} and as indicated in Fig. \ref{fig:PolesPicture}, particularly the decaying and capturing poles are affected by the centrifugal barrier. For s-wave collisions, these poles collide in the lower half of the complex momentum plane upon increasing the potential interaction strength and turn into virtual states. For higher partial waves on the other hand, the centrifugal barrier shifts the collision point to the origin. This means that the capturing state turns into a virtual state, whereas the decaying state immediately turns into a bound state. 
Physically we can regard this process as the decaying state corresponding to a quasi-bound state, or shape resonance, trapped behind the centrifugal barrier and transforming into a true bound state upon an increase in the potential depth. \par 
Since only the poles that are sufficiently close to the real axis of the complex momentum plane produce experimentally observable sudden changes in the phase shift $\delta_{\ell}(k)$, it is generally possible to truncate the infinite product over the $\mathrm{n}$ states in Eq.~\eqref{eq:Sexpansion}. 
As, contrary to s-wave scattering, the complex poles for $\ell \neq 0$ scattering processes propagate close to the real momentum axis before they collide at the origin, they generally provide a non-negligible contribution to the phase shift. 
This emphasizes the importance of the inclusion of p-wave shape resonances in the analysis of the S-matrix. 
\subsubsection{\label{subsec:Gamow states}Gamow states}
As presented in Eq.~\eqref{eq:Sexpansion}, the S-matrix can be fully expanded in terms of its poles. As such, it is convenient to introduce the set of wave functions $\Omega_{\ell,n}(r)$ that are the eigenstates of the Schr\"odinger equation with eigenvalues $k_{\mathrm{n}}$, such that 
\begin{align}
\Omega_{\ell,n}(r) = \underset{k\rightarrow k_{\mathrm{n}}}{\lim} \psi_{\ell}(k,r), 
\end{align}
which satisfy the following set of boundary conditions
\begin{align}
\Omega_{\ell,\mathrm{n}}(0) = 0 \qquad \text{and} \\
\frac{d \Omega_{\ell,\mathrm{n}}(R)}{dr} = i k_{\mathrm{n}} \Omega_{\ell,\mathrm{n}}(R),
\end{align}
where $R$ is an arbitrary distance chosen in the asymptotic regime. The wave functions $\Omega_{\ell,n}(r)$ are more commonly referred to as Gamow states, named after G. Gamow who first studied the decaying states as introduced in Sec.~\ref{subsec:Factorizing the S-matrix} in the context of alpha decay. 
Since the wave number $k_{\mathrm{n}}$ is complex for decaying states ($k =\alpha - i\beta$), the amplitude of the corresponding Gamow state grows exponentially and the solutions are non-Hermitian. This is not problematic since the decay states (as well as the virtual and capturing states) are located on the second, non-physical, energy sheet. On this sheet the Hamiltonian is not necessarily Hermitian \cite{goosen2011}.
It is however possible to form a biorthogonal set by including the dual states $\ket{\Omega_{\ell,\mathrm{n}}^D}$ that satisfy purely incoming boundary conditions and correspond to the capturing poles ($k =-\alpha - i\beta$), such that $\ket{\Omega_{\ell,\mathrm{n}}^D} \equiv \ket{\Omega_{\ell,\mathrm{n}}}^*$ and $\braket{\Omega_{\ell,\mathrm{n}}^{D}|\Omega_{\ell,\mathrm{n}'}}_R = \delta_{\mathrm{n},\mathrm{n'}}$ \footnote{Where $\braket{...}_R$ represents the regularized orthogonality condition.}. \par
In the limit where $\alpha \rightarrow 0$, the (dual) Gamow state reduces to the (virtual) bound state wave function. 
The Gamow and its dual state thus provides a useful set of eigenfunctions of the Schr\"odinger equation in the entire complex momentum plane. Their usefulness is particularly clear upon using the Mittag-Leffler therorem \cite{siegert1939} and upon realizing that the Green's function shares poles with the S-matrix, such that we can expand the Green's function in the following convergent series \cite{newton2013,kukulin2013}
\begin{widetext}
\begin{align}
\label{eq:GamowExpansion}
G^{+}_{\ell}(E,r,r') = &\sum_{\mathrm{n} = 1}^{\mathrm{N}} \frac{\Omega_{\ell,\mathrm{n}}(r)\Omega^{D,*}_{\ell,\mathrm{n}}(r')}{k_{\ell,\mathrm{n}}(k-k_{\ell,n})}+ 
\frac{1}{2} \sum_{\mathrm{n} = N+1}^{\infty} \left[\frac{\Omega_{\ell,\mathrm{n}}(r)\Omega_{\ell,\mathrm{n}}^{D,*}(r')}{k_{\ell,\mathrm{n}}(k-k_{\ell,\mathrm{n}})}-\frac{\Omega^D_{\ell,\mathrm{n}}(r)\Omega_{\ell,\mathrm{n}}^*(r')}{k^*_{\ell,\mathrm{n}}(k+k^*_{\ell,\mathrm{n}})}\right],
\end{align} 
\end{widetext}
where $\mathrm{N}$ is the (finite) number of bound and virtual states. As discussed in Sec.~\ref{subsec:Factorizing the S-matrix} it is generally possible to truncate the summation over the infinite number of complex poles to a finite number of poles with observable contributions to the phase shift.
\begin{figure*}[t!]
\centering
\includegraphics[width=\textwidth]{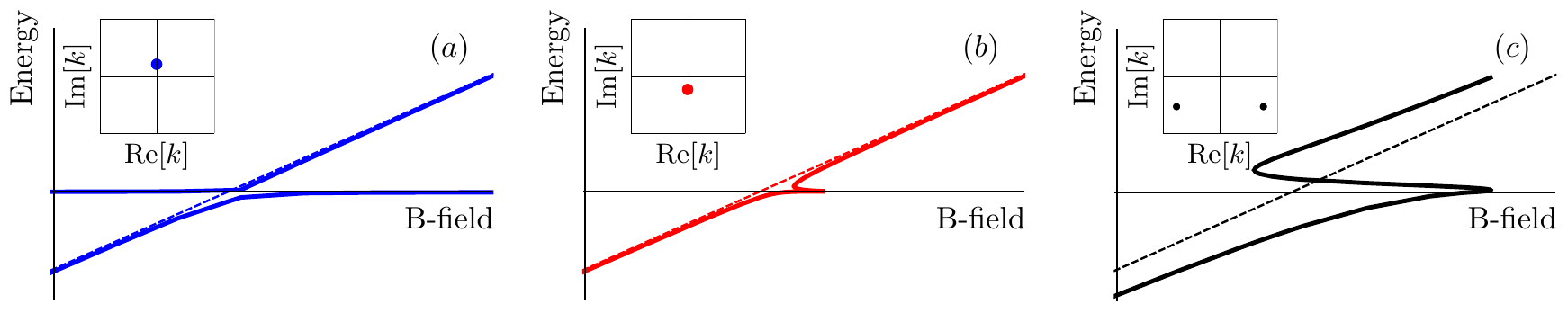}
\caption{\textbf{Energy of a Feshbach resonance versus magnetic-field} (in arbitrary units) in the presence of a near-threshold open-channel bound state (a), virtual state (b) or decaying and capturing states (c). The energy is extracted from the pole of Eq. (33) of Ref. \cite{ahmed2021}, where we use the Mittag-Leffler series presented in Eq. (36), for different values of the resonance momentum $k_n$.
The insets in all figures show the pole locations in the complex momentum plane.
\label{fig:EvsB_VariousPoles}}
\end{figure*}
\subsection{\label{subsec:Multi channel}Two-channel}
Whereas the previous section focused on single channel resonances, the multichannel nature of scattering can result in the presence of additional (near) resonant states. 
In this section we will treat the interplay between these single channel and multi channel (Feshbach) resonances. 
\subsubsection{\label{subsec:Feshbach resonances} Feshbach resonances}
For small interparticle spacings, central interactions between two particles can couple different hyperfine channels. This allows for the presence of Feshbach resonances. Contrary to single channel resonances, these Feshbach resonances are magnetic-field dependent due to the difference in the magnetic moment between hyperfine states and can hence be tuned through the variation of an externally applied magnetic-field. \par
Retaining the multichannel nature of these resonances but limiting the complexity of the analysis, these Feshbach resonances are generally treated in a two-channel model. Here, the coupling of a (near) resonant state in a closed channel subspace $\mathcal{Q}$ to the open channel subspace $\mathcal{P}$ results in the desired resonance. 
The S-matrix in the open channel subspace $\mathcal{P}$ for these two-channel models been the topic of many studies and can generally be expressed as 
\begin{align}
\label{eq:S2general}
S = S_{\mathrm{p}}\left(1-\frac{i \Gamma(E)}{E-\epsilon_c - \Delta_{\mathrm{res}}(E)+\frac{i}{2}\Gamma(E)}\right),
\end{align}
where $\Gamma(E)$ represents the resonance width, $\epsilon_c$ represents the bare resonance energy, $\Delta_{\mathrm{res}}(E)$ represents the shift of the resonance energy due to the dressing of the Feshbach state and where
$S_{\mathrm{P}}$ represents the direct open channel scattering matrix that can contain background as well as resonant effects as treated in Sec.~\ref{subsec:Factorizing the S-matrix}. \par
Short-range p-wave interactions allow for an effective range expansion in the low-energy regime. Using this expansion (stated explicitly for the phase shift in Eq.~\eqref{eq:ERApwave}), one can observe the bound-state poles located in the two-channel part of the S-matrix in Eq.~\eqref{eq:S2general} to vary linearly with magnetic-field. 
However, in the presence of (near) resonant open channel interactions, the poles in the direct part of the S-matrix $S_{\mathrm{p}}$, contained in the Ning-Hu representation of Eq.~\eqref{eq:Sexpansion}, interact with the Feshbach state and alter its magnetic-field variation.
As indicated in Figs.~\ref{fig:EvsB_VariousPoles}(a)-(b) the presence of a near-threshold open-channel bound or virtual state is only observable in a relatively small magnetic-field regime and its effect on experimental observables is hence limited. This is caused by the naturally narrow character of the p-wave interactions which is caused by the presence of the centrifugal barrier. 
On the other hand, the effect of the decaying and the capturing state in Fig.~\ref{fig:EvsB_VariousPoles}(c) has a relatively large effect compared to the bound and virtual states. These poles contain a real as well as an imaginary energy part and consequently add a width to the resonance. The apparent wide character of the resulting Feshbach state is clearly visible in Fig.~\ref{fig:EvsB_VariousPoles}(c) and is qualitatively consistent with the observed coupled-channels structure as presented in Fig.~\ref{fig:diffmu_2channel}. 
\section{\label{sec:Resonance facilitated scattering}Resonance facilitated scattering}
Whereas the two-channel model presented in Sec.~\ref{subsec:Feshbach resonances} qualitatively captures the physics observed in Fig.~\ref{fig:diffmu_2channel}, its features do not match the coupled-channels data quantitatively and one obtains non-physical values of the differential magnetic moment. In this section we investigate a three-channels model and study how its reduction to a resonance facilitated form improves on the quantitative correspondence with the CC data. 
\subsection{Full three-channel model}
We consider a three-channel system with two-atom states $\ket{bb}$ and $\ket{ac}$ coupled to a ramping Feshbach state $\ket{c}$ that consists of a (magnetic-field dependent) combination of hyperfine states. Whereas the entrance channel $\ket{bb}$ is always energetically open, the channel $\ket{ac}$, which has a threshold energy of $E_{\mathrm{th}} \approx 2.4-0.014(B-198)$ MHz in the relevant magnetic-field regime, can be either open ($E>E_{\mathrm{th}}$) or closed ($E<E_{\mathrm{th}}$). \par
The three-channel model satisfies the following Schr\"odinger equation  
\begin{align}
\label{eq:SchrodingerCoupledFull}
E\begin{pmatrix} \psi_{bb} \\ \psi_{c} \\ \psi_{ac} \end{pmatrix} = \begin{pmatrix} \hat{H}_{bb} & \hat{V}_{bb,c} & \hat{V}_{bb,ac} \\ \hat{V}_{c,bb} & \hat{H}_{c} & \hat{V}_{c,ac} \\ \hat{V}_{ac,bb} & \hat{V}_{ac,c} & \hat{H}_{ac} \end{pmatrix} \begin{pmatrix} \psi_{bb} \\ \psi_{c} \\ \psi_{ac} \end{pmatrix}, 
\end{align}
where $V_{a,b}$ represent potential operators that couple the hyperfine states and where ${H}_{a}$ is defined as $H_a= \hat{H}^0_{a}+\hat{V}_{a}$, with kinetic energy operator $\hat{H}^0_{a}$ and two-body interaction potential $\hat{V}_{a}$.
We can now proceed in two ways. First of all, we can use the operator formalism presented in App.~\ref{app:Three-channel model using the operator formalism} in order to derive an effective potential interaction $V_{\mathrm{eff}}$ that solves the Schr\"odinger equation $(E-H_{\mathrm{eff}})\ket{\psi_{bb}^+} = 0$. Alternatively, we can follow the steps presented in App.\ref{app:Details of the full three-channel model} and analyze the Lippmann-Schwinger equation for the entrance channel wavefunction $\ket{\psi_{bb}^{+}}$. Both the former as well as the latter method result in the following definition of the effective potential
\begin{align}
\label{eq:VeffFull}
V_{\mathrm{eff}} = &V_{bb,bb}+V_{bb,c}\mathrm{A}V_{c,bb}+V_{bb,c}\mathrm{A}V_{c,ac}\mathrm{G}^0_{ac}V_{ac,bb} \notag \\ 
&+V_{bb,ac}\mathrm{G}^0_{ac}V_{ac,c}\mathrm{A}V_{c,bb}+V_{bb,ac}\mathrm{G}^0_{ac}V_{ac,bb} \notag \\ 
&+V_{bb,ac}\mathrm{G}^0_{ac}V_{ac,c}\mathrm{A}V_{c,ac}\mathrm{G}^0_{ac}V_{ac,bb},
\end{align}
with propagators $G^0_{a} = (E-H_{a})^{-1}$ and with the parameter $\mathrm{A}$ defined as
\begin{align}
\label{eq:A}
\mathrm{A} = \frac{\ket{\phi_c}\bra{\phi_c}}{E-\epsilon_c}\left[1-\frac{\braket{\phi_c|\hat{V}_{c,ac}\left(E-\hat{H}_{ac}\right)^{-1}\hat{V}_{ac,c}|\phi_c}}{E-\epsilon_c}\right]^{-1},
\end{align}
where we have introduced the complex energy shift
\begin{align} 
A_{\mathrm{ac}}(E) = \braket{\phi_c|\hat{V}_{c,ac}\left(E-\hat{H}_{ac}\right)^{-1}\hat{V}_{ac,c}|\phi_c}
\end{align}
and where we have used the single resonance approximation to express the propagator $ G^0_{c}$ in the Feshbach channel $\ket{c}$ as 
\begin{align}
G^0_{c,c} = \frac{\ket{\phi_c}\bra{\phi_c}}{E-\epsilon_c}.
\end{align}
Physically, the parameter $\mathrm{A}$ indicates how the Feshbach state can either propagate freely in the $\ket{c}$ channel or couple to the $\ket{ac}$ channel where it propagates before coupling back to the $\ket{c}$ channel. 
Since $V_{\mathrm{eff}}\ket{\psi_{bb}^{+}} = T_{bb,bb}\ket{k}$ \cite{goosen2011}, the knowledge of the effective potential allows for the computation of the  entrance channel transition matrix element $T_{bb,bb}$ and hence the scattering matrix element $S_{bb,bb}$ \footnote{The T-matrix is directly related to the energy-normalized S-matrix as $\mathrm{S}= 1-2\pi i \mathrm{T}$.}. \par 
\subsection{Resonance-facilitated three-channel model}
\label{subsec:ResonanceFacilitatedThreeChannelModel}
As previously stated in the Introduction, the near-threshold behavior of the $B = 198.8$ G ($M_{L} = 0$) and $B = 198.3$ G ($M_{L} = \pm 1$) resonances is expected to be well-described in terms of a resonance-facilitated model where we neglect the direct coupling between the $\ket{bb}$ and the $\ket{ac}$ channels ($V_{bb,ac}=V_{ac,bb}=0$). \par 
This approximation simplifies the three-channel model significantly as only the first two terms in the effective potential defined in Eq.~\eqref{eq:VeffFull} remain.
We recognize that all the information regarding the coupling between the Feshbach state $\ket{c}$ and the $\ket{ac}$ channel is now contained in the single parameter $\mathrm{A}$. In order to completely isolate the effect of the $\ket{ac}$ channel on the model, we introduce the dressing factor $\mathrm{D}$ as 
\begin{align}
\label{eq:D}
\mathrm{D} &= \mathrm{A}\left(\frac{\ket{\phi_c}\bra{\phi_c}}{E-\epsilon_c}\right)^{-1}.
\end{align}
In the two-channel limit ($V_{ac,c} = V_{c,ac} = 0$), we find that $\mathrm{D}=1$ and Eq.~\eqref{eq:VeffFull} reduces to the well-known two-channel effective potential. 
Conveniently, the resonance dressing factor allows us to recast the scattering matrix element $S_{bb,bb}$ into the form
\begin{widetext}
\begin{align}
\label{eq:3channelS}
S_{bb,bb} = S_{\mathrm{P}}\left(1-\frac{2\pi i \abs{\braket{\phi_c|\hat{V}_{c,bb}|\phi_{bb}^+}}^2}{\frac{E-\delta\mu(B-B_{\mathrm{n}})}{\mathrm{D}}-\braket{\phi_c|\hat{V}_{c,bb}\left(E-\hat{H}_{bb}\right)^{-1}\hat{V}_{bb,c}|\phi_c}}\right),
\end{align}
\end{widetext}
where $\delta\mu(B-B_\mathrm{n})=\epsilon_c$, with bare magnetic resonance position $B_{\mathrm{n}}$. The combination of the complex energy shift $A_{\mathrm{bb}}(E) = \braket{\phi_c|\hat{V}_{c,bb}\left(E-\hat{H}_{bb}\right)^{-1}\hat{V}_{bb,c}|\phi_c}$ and the resonance dressing factor in the denominator of Eq.~\eqref{eq:3channelS} shift the resonance location to its dressed magnetic-field value $B_0$ and add a width to the resonance. The analysis of this shift and width is presented in Sec.~\ref{subsec:Gamow expansion for bb and a}. 
Equation \eqref{eq:3channelS} implies that the two-channel S-matrix can be used and updated to the three-channel resonance-facilitated model by a simple implementation of the resonance dressing factor.  
Physically, this factor describes how the free propagation in the Feshbach state is dressed by the $\ket{ac}$ channel. The details of this dressing and the physics of the dressing factor will be discussed in more detail in Sec.~\ref{subsec:Gamow expansion for bb and a}.
\subsection{Gamow expansion for $\ket{bb}$ and $\ket{ac}$}
\label{subsec:Gamow expansion for bb and a}
Equation \eqref{eq:3channelS} depends on the propagators $G^0_{bb} = (E-\hat{H}_{bb})^{-1}$ and $G^0_{ac} = (E-\hat{H}_{ac})^{-1}$ through the complex energy shifts $A_{\mathrm{ac}}(E)$ and $A_{\mathrm{bb}}(E)$ respectively.
As discussed in Sec.~\ref{subsec:CC structure}, both the $\ket{bb}$ as well as the $\ket{ac}$ channel has a near-threshold shape resonance. Considering that there are no other near-threshold poles in these channels, the shape resonances are the dominant contributions to the Mittag-Leffler expansion, or the Gamow expansion as presented in Eq.~\eqref{eq:GamowExpansion}, such that we can approximate the propagators $G^0_{bb}$ and $G^0_{ac}$ as 
\begin{align}
\label{eq:BBpropagator}
G^0_{bb} &= \left(\frac{\ket{\Omega_{bb}}\bra{\Omega_{bb}^D}}{2 k_{bb} (k-k_{bb})} - \frac{\ket{\Omega_{bb}^D}\bra{\Omega_{bb}}}{2 k^*_{bb} (k+k^*_{bb})}\right), \\
\label{eq:ACpropagator}
G^0_{ac} &= \left(\frac{\ket{\Omega_{ac}}\bra{\Omega_{ac}^D}}{2 k_{ac} ((k_{\mathrm{sh}}-k_{ac})} - \frac{\ket{\Omega_{ac}^D}\bra{\Omega_{ac}}}{2 k^*_{ac} (k_{\mathrm{sh}}+k^*_{ac})}\right),
\end{align}
where we have introduced the Gamow states $\ket{\Omega_{bb}}$ and $\ket{\Omega_{ac}}$ as well as their dual states $\ket{\Omega_{bb}^D} \equiv \ket{\Omega_{bb}}^*$ and $\ket{\Omega_{ac}^D} \equiv \ket{\Omega_{ac}}^*$. 
In addition, the shifted momentum $k_{\mathrm{sh}}$ in Eq.~\eqref{eq:ACpropagator} is defined as $k_{\mathrm{sh}} = \sqrt{E-E_{\mathrm{th}}}$ and accounts for the energy threshold difference between the $\ket{bb}$ and $\ket{ac}$ channels as discussed in Sec.~\ref{subsec:CC structure}. \par 
Substituting Eqs.~\eqref{eq:BBpropagator} and \eqref{eq:ACpropagator} into Eqs. \eqref{eq:3channelS} and \eqref{eq:A}, we find that \cite{ahmed2021}
\begin{align}
A_{\mathrm{bb}}(E) &\approx \frac{\braket{\phi_c|H_{c,bb}|\Omega_{bb}}\braket{\Omega_{bb}^D|H_{bb,c}|\phi_c}}{2k_{bb}(k-k_{bb})}-\notag \\ 
&\frac{\braket{\phi_c|H_{c,bb}|\Omega_{bb}^D}\braket{\Omega_{bb}|H_{bb,c}|\phi_c}}{2k_{bb}^*(k+k_{bb}^*)}
\end{align}
and 
\begin{align}
A_{\mathrm{ac}}(E) &\approx \frac{\braket{\phi_c|H_{c,ac}|\Omega_{ac}}\braket{\Omega_{ac}^D|H_{bb,c}|\phi_c}}{2k_{ac}(k_{\mathrm{sh}}-k_{ac})}- \notag \\ 
&\frac{\braket{\phi_c|H_{c,ac}|\Omega_{ac}^D}\braket{\Omega_{ac}|H_{ac,c}|\phi_c}}{2k_{ac}^*(k_{\mathrm{sh}}+k_{ac}^*)}.
\end{align}
Using the Wigner threshold scaling of the Gamow states as outlined in Ref.~\cite{ahmed2021} and using the definition 
$A(E) = \Delta_{\mathrm{res}}(E)-\frac{i}{2}\Gamma(E)$ with energy shift $\Delta_{\mathrm{res}}(E)$ and energy width $\Gamma(E)$, we can obtain
\begin{align}
\label{eq:GamowDeltaResFB}
\Delta_{\mathrm{res},bb}(E) &\approx g_{bb,c} \Re \left\{ \frac{E_{bb}^{3/2}}{E-E_{bb}} \right\} \\
\label{eq:GamowGammaFB}
\Gamma_{bb}(E) &\approx -2 g_{bb,c} \frac{E^{3/2} }{\left|E-E_{bb} \right|^2} \Im \{ E_{bb} \}, 
\end{align}
with momentum independent coupling strength $g_{bb,c}$. Whereas the $\ket{bb}$ channel is always energetically open, the $\ket{ac}$ channel can be either open ($E \geq E_{\mathrm{th}}$) or closed ($E < E_{\mathrm{th}}$). Carefully distinguishing between these two regimes we find that 
\begin{widetext}
\begin{align}
\label{eq:Deltaresac}
\Delta_{\mathrm{res},ac}(E) \approx \left\{
        \begin{array}{ll}
             g_{ac,c} \Re \left\{ \frac{E_{ac}^{3/2}}{(E-E_{\mathrm{th}})-E_{ac}} \right\}  & \quad \text{for} \, E\geq E_{\mathrm{th}} \\
            g_{ac,c}\left(\Re \left\{ \frac{E_{ac}^{3/2}}{(E-E_{\mathrm{th}})-E_{ac}} \right\}+i \frac{(E-E_{\mathrm{th}})^{3/2} }{\left|(E-E_{\mathrm{th}})-E_{ac} \right|^2} \Im \{ E_{ac} \}\right)   &  \quad \text{for} \, E< E_{\mathrm{th}} 
        \end{array}
    \right. 
\end{align}
and 
\begin{align}
\label{eq:Gammaac}
\Gamma_{ac}(E) \approx \left\{
        \begin{array}{ll}
             -2 g_{ac,c} \frac{(E-E_{\mathrm{th}})^{3/2} }{\left|(E-E_{\mathrm{th}})-E_{ac} \right|^2} \Im \{ E_{ac} \}  & \quad \text{for} \, E\geq E_{\mathrm{th}} \\
            0 & \quad \text{for} \, E< E_{\mathrm{th}} 
        \end{array}
    \right. 
\end{align}
with momentum independent coupling strength $g_{ac,c}$. Using the definitions of $A_{\mathrm{bb}}(E)$ and $A_{\mathrm{ac}}(E)$, we can rewrite the dressing factor $\mathrm{D}$ as
\begin{align}
\label{eq:Dfactorfinal}
\mathrm{D} = \frac{E-\delta\mu(B-B_{\mathrm{n}})}{E-\delta\mu(B-B_{\mathrm{n}})-\Delta_{\mathrm{res},ac}(E)+\frac{i}{2}\Gamma_{ac}(E)},
\end{align}
and we can express the S-matrix component $S_{bb,bb}$ presented in Eq.~\eqref{eq:3channelS} as
\begin{align}
\label{eq:3channelSFinal}
S_{bb,bb} = S_{\mathrm{P}}\left(1-\frac{i\Gamma_{bb}(E)}{E-\delta\mu(B-B_{\mathrm{n}})-(\Delta_{\mathrm{res},bb}(E)+\Delta_{\mathrm{res},ac}(E))+\frac{i}{2}(\Gamma_{bb}(E)+\Gamma_{ac}(E))}\right)
\end{align}
\end{widetext}
The insightful form of Eq.~\eqref{eq:3channelSFinal} reveals how the complex energy shifts set by the coupling of the Feshbach state to the $\ket{bb}$ and $\ket{ac}$ channels emerge on equal footing in the denominator of Eq.~\eqref{eq:Dfactorfinal}.  As indicated by Eq.~\eqref{eq:Gammaac}, the $\ket{ac}$ channel only contributes a finite width factor $\frac{i}{2} \Gamma_{ac}(E)$ once the channel becomes energetically open. Physically, this term in the denominator captures the resonance-facilitated loss of the $\ket{bb}$ state to the $\ket{ac}$ state. Therefore, contrary to the well-known two-channel Feshbach formalism, the presented three-channel model is capable of including inelastic loss processes, such that $S_{bb,bb}$ becomes non-unitary ($\abs{S_{bb,bb}} <1$) for $E\geq E_{\mathrm{th}}$.
\section{\label{sec:ResonanceWidth3CH} Field dependence of resonance scattering parameters}
Having factored out the dipole-dipole contribution, the low-energy scaling of the  scattering phase shift $\delta(k)$ follows the typical ERA for p-wave interactions, where 
\begin{align}
\label{eq:ERApwave}
\cot \delta(k) = -(V k^3)^{-1}-(R k)^{-1}+\mathcal{O}\{k\},
\end{align}
with scattering volume $V$ and effective range $R$. 
Using the multiplicative nature of the total S-matrix as outlined in Sec.~\ref{sec:Factorisation of the S-matrix}, the ERA can be applied to the direct part of the scattering matrix $S_{\mathrm{P}}$ and the Feshbach part $S_{\mathrm{FB}}$ separately. Analysing each of these contributions in the following two subsections, the combined scattering volume and effective range can be computed from the two partial contributions as  
\begin{align}
\label{eq:ScatteringVolumeTotal}
V = V_{\mathrm{1}}+V_{\mathrm{2}} 
\end{align}
and 
\begin{align}
\label{eq:Rtotal}
R^{-1} = \frac{V^2_1}{V^2 R_1}+ \frac{V^2_2}{V^2 R_2}.
\end{align}
The separate analysis of the contributions to the ERA will be particularly useful in the resonance width classification as presented in Sec.~\ref{subsec:Resonance width classification}.  
\subsection{\label{subsec:Direct scattering parameters} Direct scattering parameters}
Applying the parametrization of the $\ket{bb}$ channel shape resonance as presented in Sec.~\ref{subsec:Gamow expansion for bb and a}, the Ning-Hu representation of the direct part of the scattering matrix $S_{\mathrm{P}}$ is equivalent to the form presented in Ref. \cite{ahmed2021}, such that \cite{Ning:1948NingHuRepresentation,ahmed2021}
\begin{align}
\label{eq:NingHu2poles}
S_\mathrm{P} = e^{-2i k r_c}\frac{(k-k_{bb}^*)(k+k_{bb})}{(k-k_{bb})(k+k_{bb}^*)}.
\end{align}
Setting $r_c$ to equal $2 \text{Im}\{k_{bb}^{-1}\}$, we ensure that $S_{\mathrm{P}}$ follows the correct low-energy p-wave Wigner scaling.
The $k \rightarrow 0$ limit of $S_{\mathrm{P}}$ then yields the following expression for the scattering volume \cite{ahmed2021}
\begin{align}  \label{eq:VP}
V_\mathrm{P} = -r_c |k_{bb}|^{-2} \left(1 - {|r_c k_{bb}|^2}/{3} \right) 
\end{align}
and the effective range 
\begin{align} \label{eq:RP}
R_\mathrm{P} = \frac{ r_c (1-|r_c k_{bb}|^2/3)^2}{1-|r_c k_{bb}|^2+|r_c k_{bb}|^4/5}.
\end{align}
\subsection{\label{subsec:Feshbach scattering parameters} Feshbach scattering parameters}
The Feshbach contribution to the scattering volume and effective range can be directly obtained from the low-energy expansion of the Feshbach part of Eq.~\eqref{eq:3channelSFinal}. An important subtlety in this analysis is the presence of the threshold energy shift $E_{\mathrm{th}}$ in the energy width $\Gamma_{\mathrm{res},ac}(E)$ and energy shift $ \Delta_{\mathrm{res},ac}(E)$. In the absence of $E_{\mathrm{th}}$ we find that 
\begin{align}
\Delta_{\mathrm{res}}(E) &\underset{k \rightarrow 0} \approx -gk_{R}-g \frac{k_{R}}{k_{R}^2+k_{I}^2} k^2 + \mathcal{O}(k^4) \\ 
& \:= \Delta_{res}^0 + \Delta_{res}^{1} k^2 +\mathcal{O}(k^4) \notag  
\end{align}
and
\begin{align}
\Gamma(E) &\underset{k \rightarrow 0} \approx -\frac{4g k_{I} k_{R}}{(k_{I}^2+k_{R}^2)^2} k^3 + \mathcal{O}(k^5) \\ 
&\: = \Gamma^0 k^3+\Gamma^1 k^5+\mathcal{O}(k^7),
\end{align}
with $k_\mathrm{n} = k_R+i k_I$.
Using the previous expressions and isolating the effect of $E_{\mathrm{th}}$ on the $\ket{ac}$ channel parameters, we can use Eq.~\eqref{eq:3channelSFinal} in order to find the following form of the Feshbach part of the scattering volume
\begin{align}
\label{eq:Nu3chan}
V_{\mathrm{FB}}(B) = -\frac{\Gamma^0_{bb}/2}{\delta\mu(B-B_n)+\Delta_{\mathrm{res},bb}^{0}+\Delta_{\mathrm{res},ac}^{0}\chi},
\end{align}
where all dependence on $E_{\mathrm{th}}$ is contained in the parameter $\chi(E_{\mathrm{th}})$, defined as
\begin{align}
\chi = \frac{k_{I,ac}^2+k_{R,ac}^2-2 k_{I,ac} k_{\mathrm{th}}}{k_{I,ac}^2+k_{R,ac}^2+k_{th}^2-2 k_{I,ac} k_{\mathrm{th}}}
\end{align}
In the limit of a vanishing asymptotic energy difference between the channels $\ket{ac}$ and $\ket{bb}$, the shift parameter reduces to  $\chi(k_{\mathrm{th}}) \rightarrow 1$ and the energy shift $\Delta_{\mathrm{res},ac}^{0}$ has to be treated on equal footing with the direct entrance channel shift $\Delta_{\mathrm{res},bb}^0$. In the opposite limit where the energy shift becomes large, we find $\chi(k_{\mathrm{th}}) \rightarrow 0$, such that the scattering volume is insensitive to the energetically far removed channel $\ket{ac}$. \par 
Proceeding with the analysis of the Feshbach part of the effective range $R_{\mathrm{FB}}$, we find  
\begin{align}
\label{eq:Reff3chan}
R_{FB} &= \frac{(\Gamma_{bb}^0)^2}{2}\left[\Gamma_{bb}^0(1-\Delta_{\mathrm{res},bb}^1-\xi \Delta_{\mathrm{res},ac}^1)+ \right. \notag \\ 
&\left. \Gamma_{bb}^1(\delta\mu(B-B_n)+\Delta_{\mathrm{res},bb}^0+\chi \Delta_{\mathrm{res},ac}^0)\right]^{-1},
\end{align}
with 
\begin{align} 
\xi = \frac{(k_{I,ac}^2+k_{R,ac}^2)(k_{I,ac}^2+k_{R,ac}^2-k_{I,ac}k_{\mathrm{th}})}{(k_{I,ac}^2+k_{R,ac}^2+k_{th}^2-2 k_{I,ac} k_{\mathrm{th}})^2}
\end{align}
Similar to the shift parameter $\chi(E_{\mathrm{th}})$, the shift parameter $\xi(E_{\mathrm{th}})$ reduces to a value of $\xi(E_{\mathrm{th}}) \rightarrow 1$ in the limit of a vanishing asymptotic energy shift and a value of $\xi(E_{\mathrm{th}}) \rightarrow 0$ in the opposite limit of large energy shifts. The values of the shift parameters in the $^{40}\mathrm{K}$ analysis will be presented in Sec.~\ref{subsec:Resonance facilitation factor results}.
\section{\label{sec:Results}Results}
\begin{figure}[H]
    \begin{center}
\includegraphics[width=0.82\columnwidth]{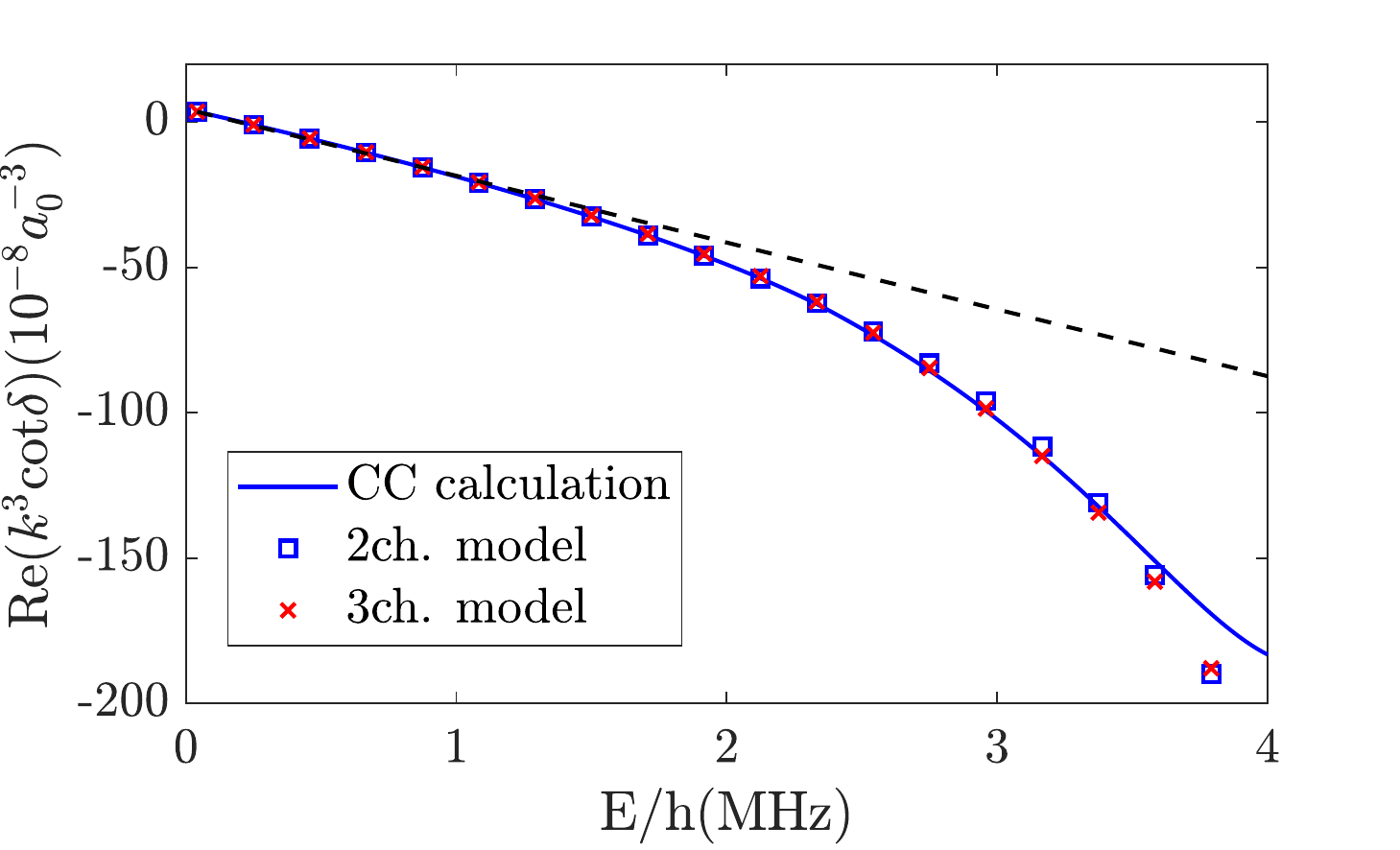}
       \caption{{\bf Scattering phase } for $M_L=0$ at 200\,G. 
The phase shift $\delta$ from CC calculations is plotted as the real part of $k^{3} \cot \delta$ (blue line) versus collisional energy $E/h$. Both the artificial two-channel (blue squares) and the realistic resonance-facilitated three-channel model (red crosses) match CC data at low energies. The black dashed curve represents the effective range expansion up to $\mathcal{O}(k^4)$.} \label{fig:PhaseShift}
\end{center}
    \end{figure}

We proceed our analysis by fitting the resonance-facilitated form of the open channel S-matrix component $S_{bb,bb}$ as presented in Eq.~\eqref{eq:3channelSFinal} to CC data in the low-energy limit. In our model, only the coupling parameters $g_{bb,c}$ and $g_{ac,c}$ are magnetic-field dependent \cite{ahmed2021}. 
Hence; we fit the shape resonance momenta $k_{bb}$ and $k_{ac}$ at a single magnetic-field value $B=200$ G and keep the best-fit values as presented in Tab. \ref{tab:ShapeResPars} fixed for all subsequent fitting routines. 
\begin{table}[H]
 \centering
 \begin{tabular}{|c|c|}
 \hline
\multicolumn{2}{|c|}{\textbf{$\ket{bb}$-channel shape resonance}}\\
 \hline
 $\Re E_{bb} $ &  $0.222 \, \bar{E}$ \\
 $\Im E_{bb} /2$ & $-0.114 \, \bar{E}$ \\
 $V_P$ & $-3.02 \, r_\mathrm{vdW}^3$ \\
 $R_P$ & $ 1.81 \,  r_\mathrm{vdW}$ \\
 \hline
 \multicolumn{2}{|c|}{\textbf{$\ket{ac}$-channel shape resonance}}\\
 \hline
  $\Re E_{ac} $ &  $0.179 \, \bar{E}$ \\
 $\Im E_{ac} /2$ & $-0.061 \, \bar{E}$ \\
 \hline
\end{tabular}
 \caption{{\bf Shape resonance parameterization.} \label{tab:ShapeResPars} The best-fit complex energies for the shape resonances in the $\ket{bb}$ and $\ket{ac}$ channel in terms of $E_{vdW}$. The scattering volume $V_\mathrm{P}$ and effective range $R_{\mathrm{P}}$ appearing introduced in Sec.~\ref{subsec:Direct scattering parameters} are also given in van der Waals units. The bare magnetic resonance position $B_{\mathrm{n}} = 168.0 \, \textrm{G}$ is set by comparison to the CC data.
\label{tab:model}}
\end{table} 
By choosing a B-field close to the resonance value, we ensure that the S-matrix exhibits rapid variations at low-energies, allowing for a well-determined fitting routine. 
To ensure physically realistic values for the fitting parameters below resonance where the S-matrix varies minimally, we force the pole of  $S_{bb,bb}$ to be located at the B-field dependent value of the binding energy $E_b$ as extracted from the CC code. This procedure directly fixes one of the free parameters $g_{bb,c}$ or $g_{ac,c}$. 
As presented in Fig. \ref{fig:PhaseShift}, the outlined fitting routine is able to correctly reproduce the CC phase shift in the low-energy regime. 
    
Whereas the two-channel fit as introduced in Ref. \cite{ahmed2021} and presented in Fig. \ref{fig:PhaseShift} similarly captures the CC data, this model relies on the use of an artificial fitting parameter and is hence less physical. Both models start to deviate from the CC data at higher scattering energies. This is a consequence of the low-energy approximations used in the Gamow expansion and the Ning-Hu expansion of the S-matrix. \par
\begin{figure}[H]
\centering
\includegraphics[width=\columnwidth]{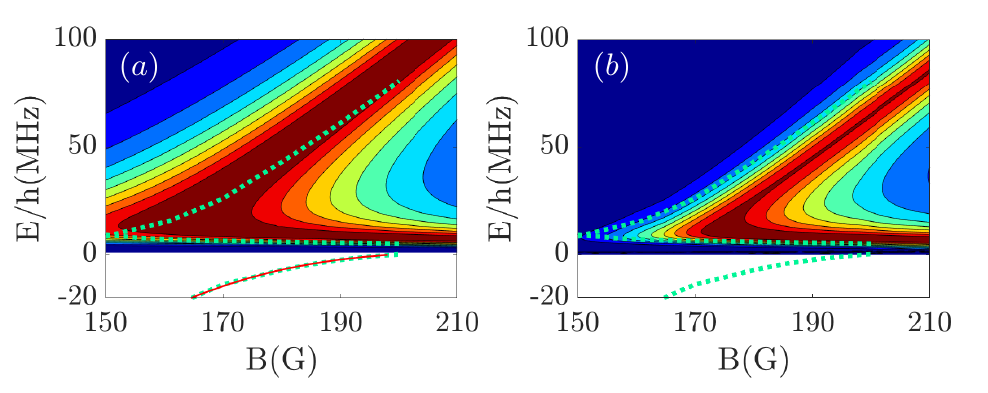}
\caption{Above threshold loss $1-\abs{S_{\mathrm{bb,bb}}}^2$ and binding energy of the $\ket{bb}$ channel computed from the CC code (left) and from the 3-channel model (right). The green dashed line corresponds to the energy extracted from the pole of the 3-channel S-matrix. In the low-energy limit, this energy approaches the quasi-bound state energy \cite{ahmed2021}. The curve has been added to both figures to ease the comparison.} 
\label{fig:ContourResults}
\end{figure}

   
The high-energy deviations are also visible in the atom loss $1-\abs{S_{\mathrm{bb,bb}}}^2$ presented in Fig. \ref{fig:ContourResults}. However, notably, the resonance-facilitated model does correctly capture the avoided crossing structure of the bound-state (and quasi-bound state) around resonance. Consistent with the CC data, the loss magnitude approaches unity once the $\ket{ac}$ channel opens and reflects the strong B-field dependent nature of the loss rate resulting from the large magnetic moment difference between the Feshbach channel on the one hand and the $\ket{bb}$ and $\ket{ac}$ channels on the other hand. 
\subsection{\label{subsec:Resonance facilitation factor results} Resonance facilitated contributions and scattering parameters}
\begin{figure}[H]
\begin{center}
\includegraphics[width=0.9\columnwidth]{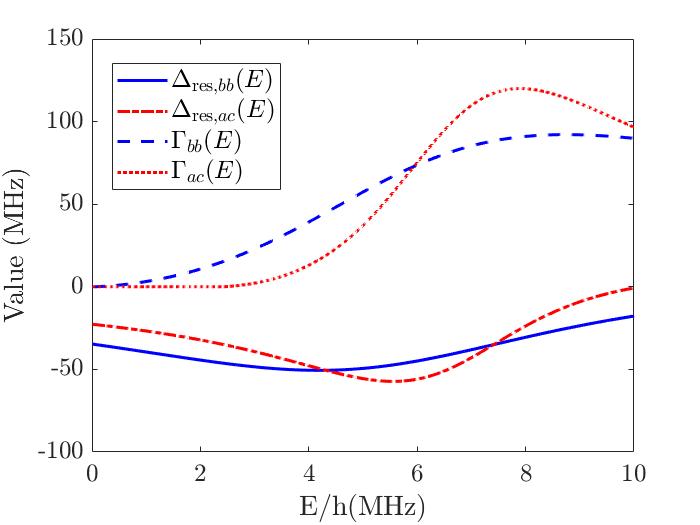}
\caption{\textbf{The energy shift and width} of the $\ket{bb}$-channel (blue lines) and the $\ket{ac}$ channel (red lines) as a function of the energy E/h at a  magnetic-field value $\mathrm{B} = 200$ G.}\label{fig:ShiftsANDWidth} 
\end{center}
\end{figure}

As indicated by the denominator of Eq.~\eqref{eq:3channelSFinal}, the dressing effects of both the $\ket{bb}$ and the $\ket{ac}$ channel on the Feshbach state arise equivalently. Both channels contribute an energy shift and, once the channels are energetically open, add a width $\Gamma(E)$. As can be seen in Fig. \ref{fig:ShiftsANDWidth}, the value of these contributions depends on the scattering energy \footnote{The parameters also depend on the B-field, but the overall observed trend holds for all considered B-fields.}. \par
\begin{figure}[H]
\begin{center}
\includegraphics[width=0.9\columnwidth]{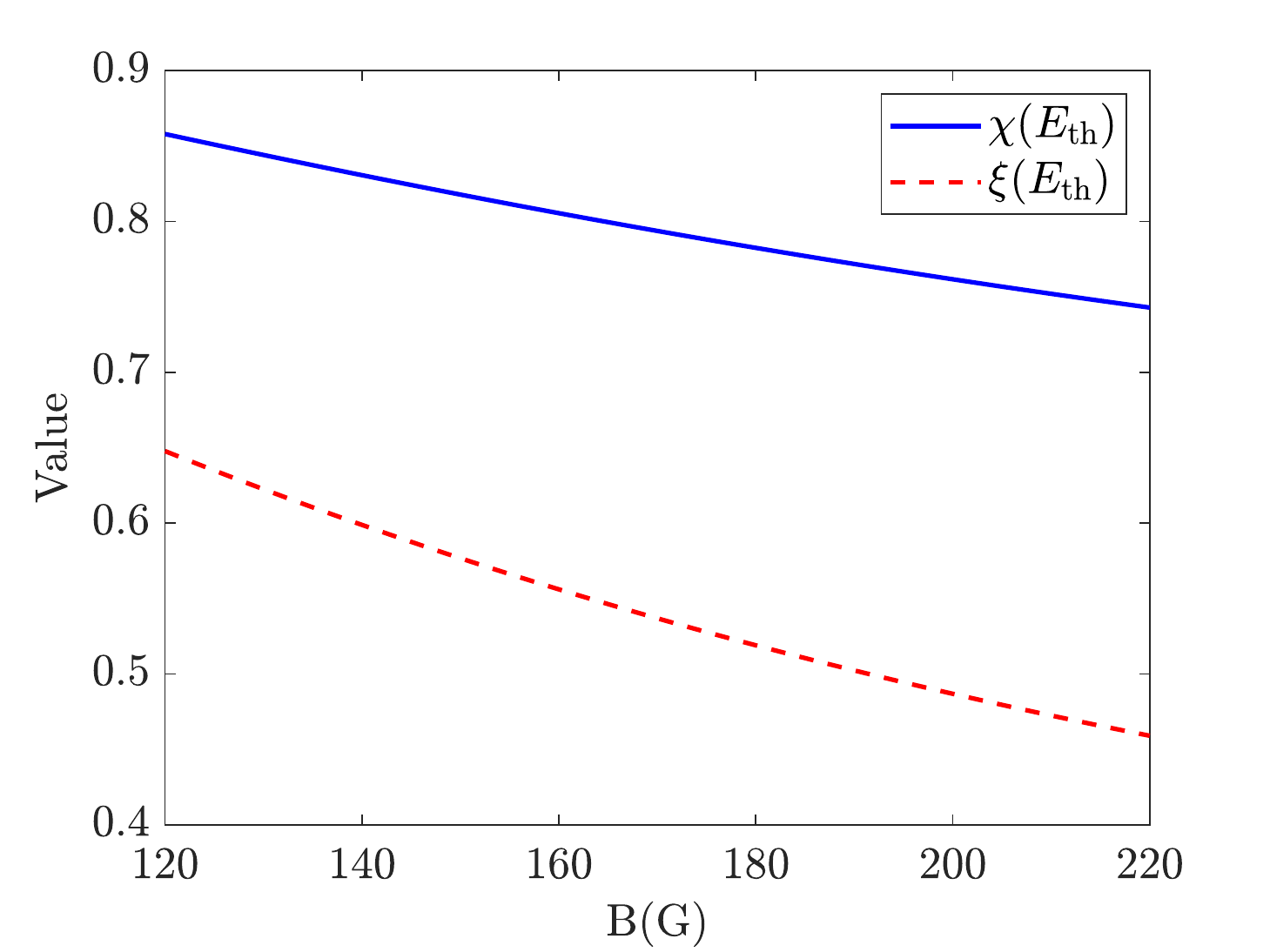}
\caption{\textbf{Resonance shift parameters.} The values of the shift parameters $\chi(E_{\mathrm{th}})$ (blue) and $\xi(E_{\mathrm{th}})$ as a function of magnetic-field.} \label{fig:chiANDxi}
\end{center}
\end{figure}

Predictably, the magnitude of the energy shift is largest when the scattering energy approaches the (real part of the) energy of the shape resonance \footnote{For the $\ket{ac}$ channel, the threshold energy has to be added since the energy is measured with respect to the $\ket{bb}$ threshold. Hence, the shift is the largest for Re($E_{bb}$)$+ E_{\mathrm{th}}$.}. In addition, the asymmetric and broad nature of the resonance widths can be contributed to the shape resonances being located above the centrifugal barrier, as indicated by the large value of the imaginary parts of the shape resonance energies.  
The comparable magnitude of the complex energy shift in the $\ket{bb}$ and $\ket{ac}$ channels implies the importance of the $\ket{ac}$ channel added in the resonance facilitated model. Considering the effective range expansion of the phase shift $\delta(k)$, it is furthermore possible to quantify the effect of the $\ket{ac}$ channel on the low-energy scattering parameters. Following the approach presented in Sec.~\ref{subsec:ResonanceFacilitatedThreeChannelModel}, we begin this analysis by computing the shift parameters $\chi(E_{\mathrm{th}})$ and $\xi(E_{\mathrm{th}})$ and obtain the results as presented in Fig. \ref{fig:chiANDxi}. 
Consistent with the analysis of the full complex energy shift, the non-negligible values of the shift parameters $\xi(E_{\mathrm{th}})$ and particularly $\chi(E_{\mathrm{th}})$ indicate the importance of the $\ket{ac}$ channel on the low-energy scattering. Whereas the values remain non-neglegible over the entire probed B-field regime, the magnitudes decrease for larger B-fields. This trend is consistent with the growing threshold energy difference between the $\ket{bb}$ and $\ket{ac}$ channels for increasing B-fields where, for larger energy differences, the  effects of the $\ket{ac}$ channel on the scattering states in the $\ket{bb}$ channel gradually becomes less important. \par
\begin{figure}[H]
\begin{center}
\includegraphics[width=0.9\columnwidth]{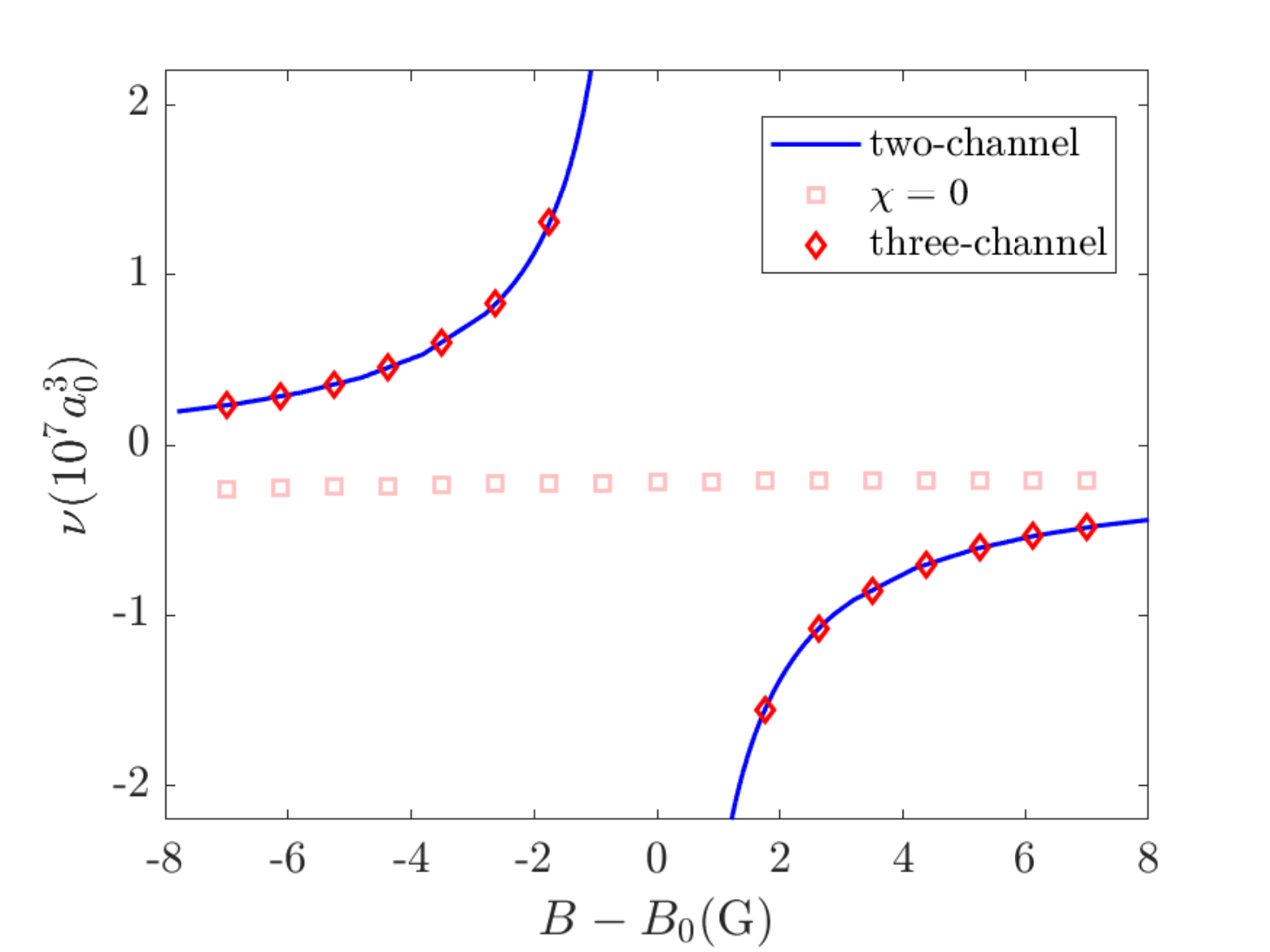}
\caption{\textbf{Scattering volume}. Artificial two-channel (blue) and three-channel (red diamond) scattering volume for $M_{\mathrm{L}} = 0$ as a function of the magnetic field. The scattering volume diverges at $B_{0} = 198.803 \text{G}$. The square data points are obtained by setting $\chi =0$ in Eq.~\eqref{eq:Nu3chan} and reveal the importance of the $\ket{ac}$ contribution to the scattering volume.} \label{fig:ERA3ch}
\end{center}
\end{figure}

The contribution of the $\ket{ac}$ channel shift parameters $\chi(E_{\mathrm{th}})$ and $\xi(E_{\mathrm{th}})$ on the $\ket{bb}$ channel scattering states is readily observed in Figs.~\ref{fig:ERA3ch} and ~\ref{fig:EffectiveRange} . Here, the computed scattering volume and the effective range are compared to the artificial two-channel model results of Ref. \cite{ahmed2021}, both in the presence of the shift parameters and in the absence of the shift parameters. 
As can be seen in Fig.~\ref{fig:ERA3ch}, the factor $\chi \Delta^0_{\mathrm{res},ac}$ has a significant impact on the computation of the scattering volume. Without this shift factor induced by the $\ket{ac}$ channel, the model is completely incapable of reproducing the CC resonance. The large impact of this factor precisely indicates why the true two-channel Feshbach model is unable to match the CC calculations and explains why in Ref. \cite{ahmed2021} the Breit-Wigner structure of this model was replaced by the artificial form presented in Eq.~\eqref{eq:2artificial}. This artificial model effectively allows for the decoupling of the fitting of the resonance widths and shifts; thereby correctly describing the large resonance shift, but consequently failing to partially attribute this shift to the $\ket{ac}$ channel and instead resulting in physically unrealistic values of $\delta\mu$. The resonance-facilitated model rectifies this physical inconsistency. 
\subsection{\label{subsec:Resonance width classification} Resonance width classification}
Whereas recent studies have revealed the universal behavior of low-energy p-wave scattering parameters, \cite{Luciuk2016,yoshida2015,yu2015,he2016,peng2016}, the multichannel nature of Feshbach resonances affects these scaling laws. The degree to which the universal behavior is distorted by multichannel effects is quantified by the resonance width.   
In this classification scheme, the magnitude of the energy-dependent contributions of the Feshbach part to the S-matrix are compared to the universal single-channel contributions. \par
Since the resonance-facilitated S-matrix retains the typical two-channel p-wave Breit-Wigner form, much of the resonance width analysis presented in Ref. \cite{ahmed2021} can be directly applied to our current model. As such, we define the 
dimensionless resonance width parameter $\zeta$ as 
\begin{align}
\label{eq:ResWidthR}
R^{-1} = \frac{-R_{\mathrm{max}}(B)^{-1}}{\zeta} \left(1-\frac{V_{\mathrm{bg}}}{V(B)}\right)^2 + R^{-1}_{\mathrm{vdW}}(B).
\end{align}
Here, $V(B)$ represents the universal Feshbach form of the scattering volume set by 
\begin{align}
V(B) = V_{\mathrm{bg}}\left(1-\frac{\Delta}{B-B_0}\right),
\end{align}
where the background scattering volume $V_{\mathrm{bg}}$ and resonance width $\Delta$ are set by the values presented in Tab. II of Ref. \cite{ahmed2021}. 
The direct single-channel contribution to the effective range in Eq.~\eqref{eq:ResWidthR} is set by the van der Waals effective range $R_{\mathrm{vdW}}$ given by 
\begin{align}
R_{\mathrm{vdW}} = R_{\mathrm{max}}\left(1+2 \frac{\bar{V}}{V(B)} + 2\frac{\bar{V}^2}{V(B)^2}\right),
\end{align}
with $R_{\mathrm{max}} \approx 76 a_0$ and $\bar{V} \approx (63.464 a_0)^3$ for two  $^{40}\mathrm{K}$ atoms. 
For broad resonances, where $\abs{\zeta} \gg 1$, the effective range is fully determined by the van der Waals contribution $R_{\mathrm{vdW}}$. On the other hand, for narrow resonances where $\abs{\zeta} \ll 1$, the effective range is determined by the Feshbach term, such that 
\begin{align}
\label{eq:Rnarrow}
R \approx R_{\mathrm{FB},0}\left(1-\frac{V_{\mathrm{bg}}}{V(B)}\right)^2,
\end{align}
with $R_{\mathrm{FB},0}$ the Feshbach part of the effective range on resonance. \par 
It is important to note that $R_{\mathrm{vdW}}$ is not identical to $R_{\mathrm{P}}$. Whereas $R_{\mathrm{vdW}}$ is a real single-channel effective range, the direct part of the effective range set by $R_{P}$ follows from CC calculations, where the boundary conditions on the $\ket{bb}$ channel are affected by the presence of coupled-channels \footnote{It is generally not true that $R_{\mathrm{P}}$ can be computed by uncoupling the channels in the CC code. The presence of other channels can significantly alter the open-channel potential structure.}. Ensuring Eq.~\eqref{eq:ResWidthR} to accurately represent the effective range of the resonance facilitated model and comparing Eqs.~\eqref{eq:Rtotal} and \eqref{eq:Rnarrow} implies that $\zeta = -R_{\mathrm{max}}/R_{\mathrm{FB},0}$. 
Here, the resonant value of Eq.~\eqref{eq:Rtotal} is fully determined by the resonant value of Eq.~\eqref{eq:Reff3chan}, which we name $R_0$. This differs from $R_{\mathrm{FB},0}$ by the resonant contribution of $R_{\mathrm{vdW}}$, set by $R_{\mathrm{max}}$. As such, we find that $R_{FB,0}^{-1} = (R_0-R_{\textrm{max}})^{-1}$ and we can directly compute the resonance parameter $\zeta$ as 
\begin{align}
\zeta = \left. \frac{\Gamma^0_{bb}}{\Gamma^0_{bb}-2 R_{\mathrm{max}}(1-\Delta_{\mathrm{res},bb}^1-\xi \Delta_{\mathrm{res},ac})} \right \rvert_{B=B_0}.
\end{align}
In agreement with Ref.~\cite{ahmed2021},  we find $\zeta = \{-1.90,-1.85 \}$ for $M_L = 0$ and $M_L = \abs{1}$ respectively. 
\begin{figure}[H]
\begin{center}
\includegraphics[width=\columnwidth]{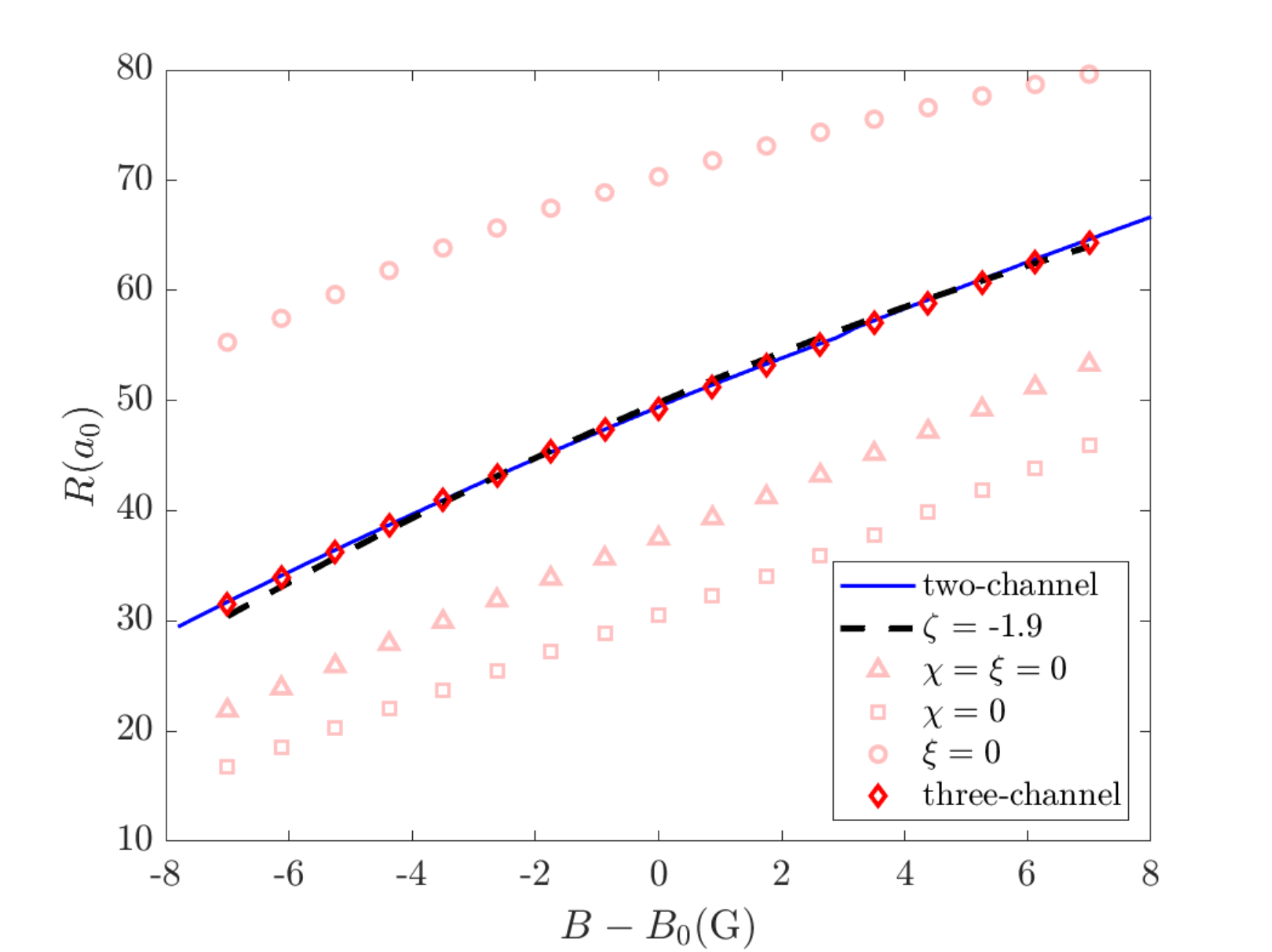}
\caption{\textbf{Effective range}. Artificial two-channel (blue) and three-channel (red diamonds) effective range for $M_{\mathrm{L}} = 0$ as a function of the magnetic field. The black dashed line represents the effective range computed using Eq.~\eqref{eq:ResWidthR} for a resonance width parameter value $\zeta = -1.9$. The square, triangular and circular data points are obtained by setting the various shift parameters to a value of zero.}  \label{fig:EffectiveRange}
\end{center}
\end{figure}
Whereas the resonance width is unaltered with respect to the two-channel model, it is important to point out that, contrary to the analysis of Ref.~\cite{ahmed2021}, the classification presented here does not rely of the fitting of $\zeta$ and instead follows directly from the three-channel model. As presented in Fig.~\ref{fig:EffectiveRange}, Eq.~\eqref{eq:ResWidthR} correctly represents the CC effective range around resonance and requires the inclusion of both $\Delta_{\mathrm{res},bb}^1$ and $\xi\Delta_{\mathrm{res},ac}^1$ to the computation of the effective range. This is in contrast with typical systems with non-resonant background interactions, where the resonance shift is well-represented by its lowest energy contribution $\Delta_{\mathrm{res}}^0$ \cite{book:chapterservaas}.  
\section{\label{sec:conclusion}Conclusion}
In this work, we upgrade the standard two-channel Feshbach formalism to a resonance-facilitated three-channel version. 
Here, the direct interaction between the open-channel and the added (third) channel is neglected and the model retains the typical Breit-Wigner form of the S-matrix. This allows for the intuitive interpretation of the results and enables the definition of the Feshbach resonance width analogous to the two-channel classification \cite{gao2011analytic}. \par
For the analyzed p-wave resonances in $^{40}\mathrm{K}$, the resonance-facilitated structure is motivated by the CC data, where a large magnetic-field dependence in the inelastic above-threshold ($E\geq 0)$ loss can be observed. The small magnetic moment difference between the open-channel and the third channel cannot account for this large dependence. As such, the CC data implies the dominant scattering processes arise through coupling to the Feshbach resonance. 
The formulated resonance-facilitated model 
 successfully captures the ``bending'' of the dimer binding energy as a function of magnetic field for the considered $^{40}\mathrm{K}$ resonances. Contrary to the two nonphysical parameters that we identified as a major shortcoming of the two-channel treatment in Ref.~\cite{ahmed2021}, the correct fitting of the resonance-facilitated model to the CC data does not require the use of nonphysical input parameters. Instead, the three-channel model presented here allows for the use of physically realistic input values. We attribute the success of the resonance-facilitated model over a  two-channel version to the explicit inclusion of a shape resonance in the added (third) channel. Similarly to the open-channel shape resonance, the interplay of this feature with the Feshbach resonance alters the low-energy scattering physics and needs to be incorporated into the model explicitly. \par 
The general framework of the resonance-facilitated three-channel model can be readily adapted to the study of other resonances \cite{https://doi.org/10.48550/arxiv.2301.08097} and systems with arbitrary partial wave interactions by the reconsideration of the low-energy scaling of the Gamow functions. Particularly systems where resonant features exist in coupled channels with similar spin-structure (singlet vs. triplet) are expected to be suitably treated by the presented model, since the  coupling strength of the direct spin-exchange interaction between these channels is expected to be small compared to channels with different spin-structure. \par 
Apart from considering interactions with different partial waves, interesting routes for further analyses include the consideration of more terms in the Ning-Hu expansion or the inclusion of higher-order $k_{s}$ terms in the Gamow series in order to improve the correctness and validity range of the model. \par 
Notably, the need for analytical three-channel models also extends to the three-body sector. 
For example, $^7\mathrm{Li}$ bosons have been observed to have a $\ket{bb}$ + Feshbach + $\ket{ac}$ s-wave structure analogous to that of the $^{40}\mathrm{K}$ p-waves developed here, and Ref \cite{https://doi.org/10.48550/arxiv.2301.08097} shows how three-channel two-body interactions modify the three-body recombination of $^7\mathrm{Li}$ atoms.  Overlapping resonances offer another example of an intrinsic three-channel system  where two different ramping closed channels produce two resonance in a single open channel.  Ref.~\cite{PhysRevA.88.052701} successfully treated such cases with an analytic multichannel quantum defect model.  It is known that a three-channel model (with one open and two closed channels) of overlapping resonances in the two-body sector is needed in order to represent correctly the experimental three-body Efimov features for three Cs atoms near a magnetic field of $B= 550$ G, where a numerically-implemented three-channel model predicted features that agree with experiment while a conventional two-channel resonance model failed to agree \cite{wang2014universal}.  Since overlapping resonances are common in many alkali and mixed-alkali systems, multichannel resonance models that go beyond the usual two-channel isolated Feshbach picture represent a promising area for future research

\section{\label{sec:acknowledgements}Acknowledgments}
The authors thank K.G Jackson, J.H. Thywissen, J. van de Kraats and J.-L. Li for useful discussions. This work is supported by the Netherlands Organisation for Scientific Research (NWO) under grant 680-47-623.
\bibliographystyle{apsrev4-1}
\bibliography{biblio_3channel}
\appendix
\section{Three-channel model using the operator formalism} \label{app:Three-channel model using the operator formalism}
We now aim to obtain the effective potential $V_{\mathrm{eff}}$ as presented in Eq.~\eqref{eq:VeffFull} using the operator formalism. 
Our starting point is the relation between the Green's function $\mathrm{G}$ and the transition operator $\mathrm{T}$, which reads
\begin{align}
\mathrm{T} = \mathrm{V}+\mathrm{VGV}.
\end{align}
Considering the open-channel component $\mathrm{T}_{bb,bb}$, the above matrix equation reduces to 
\begin{align}
\label{eq:Tbbop}
\mathrm{T}_{bb,bb} &= \mathrm{V}_{bb,bb}+\mathrm{V}_{bb,bb}\mathrm{G}_{bb,bb}\mathrm{V}_{bb,bb}+\mathrm{V}_{bb,bb}\mathrm{G}_{bb,c}\mathrm{V}_{c,bb}+ \notag \\ 
& \mathrm{V}_{bb,c}\mathrm{G}_{c,bb}\mathrm{V}_{bb,bb}+\mathrm{V}_{bb,c}\mathrm{G}_{c,c}\mathrm{V}_{c,bb}.
\end{align}
In order to get rid of the Green's functions $\mathrm{G}_{bb,c}$, $\mathrm{G}_{c,bb}$ and $\mathrm{G}_{c,c}$ we use the definition of the Green's function $\mathrm{G}(E-H) \equiv 1$, finding the following set of useful relations
\begin{align}
\label{eq:Gbbbb}
\mathrm{G}_{bb,bb} &= \mathrm{G}^0_{bb}-\mathrm{G}_{bb,c}\mathrm{H}_{c,bb}\mathrm{G}^0_{bb} \\
\label{eq:Gbbc}
\mathrm{G}_{bb,c} &= -\mathrm{G}_{bb}\mathrm{H}_{bb,c}\mathrm{G}^0_{c}-\mathrm{G}_{bb,ac}\mathrm{H}_{ac,c}\mathrm{G}^0_{c} \\
\label{eq:Gbbac}
\mathrm{G}_{bb,ac} &= - \mathrm{G}_{bb,c}\mathrm{H}_{c,ac}\mathrm{G}^0_{ac} \\ 
\label{eq:Gcc}
\mathrm{G}_{c,c} &= \mathrm{G}^0_{c}-\mathrm{G}_{c,ac}\mathrm{H}_{ac,c}\mathrm{G}^0_{c}-\mathrm{G}_{c,bb}\mathrm{H}_{bb,c}\mathrm{G}^0_{c} \\
\label{eq:Gcac}
\mathrm{G}_{c,ac} &= -\mathrm{G}_{c,c}\mathrm{H}_{c,ac}\mathrm{G}^0_{ac},
\end{align}
where we have introduced the single-channel Green's functions $\mathrm{G}^0_{p,q}$ defined as 
\begin{align}
\label{eq:Gsingle}
\mathrm{G}^0_{p,q} = (E-\mathrm{H}_{p,q})^{-1}
\end{align}
Substituting Eq.~\eqref{eq:Gbbac} into Eq.~\eqref{eq:Gbbc}, we find that we can express $\mathrm{G}_{bb,c}$ as 
\begin{align}
\label{eq:Gbbc1}
\mathrm{G}_{bb,c} = -\mathrm{G}_{bb,bb}\mathrm{H}_{bb,c}\mathrm{G}^0_{c}\left[1-\mathrm{H}_{c,ac}\mathrm{G}^0_{ac}\mathrm{H}_{ac,c}\mathrm{G}^0_{c}\right]^{-1}
\end{align}
For notational convenience, we introduce the following parameter
\begin{align}
\label{eq:Wbb}
\mathrm{W}_{bb,bb} = \mathrm{H}_{bb,c}\mathrm{G}^0_{c}\mathrm{H}_{c,bb}\left[1-\mathrm{H}_{c,ac}\mathrm{G}^0_{ac}\mathrm{H}_{ac,c}\mathrm{G}^0_{c}\right]^{-1}.
\end{align}
Since $\mathrm{G}_{bb,c} \mathrm{V}_{c,bb} = \mathrm{V}_{bb,c}\mathrm{G}_{c,bb}$, with $\mathrm{H}_{p,q} = -\mathrm{V}_{p,q}$ for all $p \neq q$, we can then obtain 
\begin{align}
\label{eq:GbbcVcbb}
\mathrm{G}_{bb,c} \mathrm{V}_{c,bb} = \mathrm{V}_{bb,c}\mathrm{G}_{c,bb} = \mathrm{G}_{bb,bb} \mathrm{W}_{bb,bb}
\end{align}
If we wish to also get rid of $\mathrm{G}_{c,c}$ and $\mathrm{G}_{c,ac}$, we can substitute Eq.~\eqref{eq:Gcac} into Eq.~\eqref{eq:Gcc}, such that 
\begin{align}
\label{eq:Gcc2}
\mathrm{G}_{c,c} = \mathrm{G}^0_{c} +\mathrm{G}_{c,c}\mathrm{H}_{c,ac}\mathrm{G}^0_{ac} \mathrm{H}_{ac,c}\mathrm{G}^0_{c}-\mathrm{G}_{c,bb}H_{bb,c}\mathrm{G}^0_{c},
\end{align}
which can be rewritten into a form from which we can extract $\mathrm{G}_{c,c}$, such that 
\begin{align}
\label{eq:Gcc3}
\mathrm{G}_{c,c} = \frac{\mathrm{G}^0_{c}-\mathrm{G}_{c,bb}\mathrm{H}_{bb,c}\mathrm{G}^0_{c}}{1-\mathrm{H}_{c,ac}\mathrm{G}^0_{ac}\mathrm{H}_{ac,c}\mathrm{G}^0_{c}}
\end{align}
Using Eqs. \eqref{eq:Gcc3} and \eqref{eq:Gbbbb}, we can now rewrite the transition operator element $\mathrm{T}_{bb,bb}$ as defined in Eq.~\eqref{eq:Tbbop} into the following form 
\begin{align}
\mathrm{T}_{bb} = &\mathrm{V}_{bb,bb}+\mathrm{V}_{bb,bb}\mathrm{G}_{bb,bb}\mathrm{V}_{bb,bb}+\mathrm{V}_{bb,bb}\mathrm{G}_{bb,bb}\mathrm{W}_{bb,bb}+\notag \\ 
&\mathrm{G}_{bb,bb}\mathrm{W}_{bb,bb}\mathrm{V}_{bb,bb}+ \mathrm{V}_{bb,c}\mathrm{G}^0_{c}\left[1-\mathrm{H}_{c,ac}\mathrm{G}^0_{ac}\mathrm{H}_{ac,c}\mathrm{G}^0_{c}\right]^{-1}\notag \\ 
&\mathrm{V}_{c,bb}-\mathrm{V}_{bb,c}\mathrm{G}_{bb,bb}\mathrm{W}_{bb,bb}
\end{align}
We can now get rid of $\mathrm{G}_{bb,bb}$ by substituting Eq.~\eqref{eq:GbbcVcbb} into Eq.~\eqref{eq:Gbbbb}, such that we can obtain the T-matrix as presented in Eq. (B11) of ref. \cite{MultichannelThomas} with $\mathrm{W}_{bb,bb} = \mathrm{W}_{bb,bb}^{\mathrm{2ch}} \cdot \mathrm{D}$. 
\begin{widetext}
\section{Details of the full three-channel model}
\label{app:Details of the full three-channel model}
In this appendix, we derive the full three-channel version of the transition matrix, indicating its comparative complexity to the resonance-facilitated version. 
For the full three-channel model, we have the following set of Lipmann-Schwinger equations 
\begin{align}
\label{eq:LSbbfull}
&\ket{\psi_{bb}^+} = \ket{\phi_{bb}^+}+\mathrm{G}^0_{bb}\left[\hat{V}_{bb,c}\ket{\psi_{c}}+\hat{V}_{bb,ac}\ket{\psi_{ac}}\right], 
\end{align}
where
\begin{align}
&\ket{\phi_{bb}^+} = \ket{\chi}+\frac{1}{E-\hat{H}_{bb}}\hat{V}_{bb,bb}\ket{k},
\end{align}
with unscattered wavefunction $\ket{\chi}$ and
\begin{align}
\label{eq:LSrrfull}
&\ket{\psi_{c}} = \mathrm{G}^0_{c}\left[\hat{V}_{c,bb}\ket{\psi_{bb}^+}+\hat{V}_{c,ac}\ket{\psi_{ac}}\right], \\
\label{eq:LSacfull}
&\ket{\psi_{ac}} = \mathrm{G}^0_{ac}\left[ \hat{V}_{ac,c}\ket{\psi_{c}}+\hat{V}_{ac,bb}\ket{\psi_{bb}}\right].
\end{align}
We aim to eliminate the wave functions $\ket{\psi_{c}}$ and $\ket{\psi_{ac}}$ from the previous set of expressions. Starting with the substitution of Eq.~\eqref{eq:LSacfull} into Eqs. \eqref{eq:LSbbfull} and \eqref{eq:LSrrfull} we find 
\begin{align}
\label{eq:LSbbfull2}
&\ket{\psi_{bb}^+} = \ket{\phi_{bb}^+}+\mathrm{G}^0_{bb}\left[\hat{V}_{bb,c}\ket{\psi_{c}}+\hat{V}_{bb,ac}\mathrm{G}^0_{ac} \left[ \hat{V}_{ac,c}\ket{\psi_{c}}+\hat{V}_{ac,bb}\ket{\psi_{bb}}\right]\right] \\
\label{eq:LSrrfull2}
&\ket{\psi_{c}} = \mathrm{G}^0_{c}\left[\hat{V}_{c,bb}\ket{\psi_{bb}^+}+\hat{V}_{c,ac}\mathrm{G}^0_{ac} \left[ \hat{V}_{ac,c}\ket{\psi_{c}}+\hat{V}_{ac,bb}\ket{\psi_{bb}}\right]\right] 
\end{align}
Next, we rewrite Eq.~\eqref{eq:LSrrfull2} in order to find the following expression for $\ket{\psi_{c}}$
\begin{align}
\ket{\psi_{c}} = \mathrm{A} \left[\hat{V}_{c,bb}\ket{\psi_{bb}^+}+\hat{V}_{c,ac}\mathrm{G}^0_{ac}\hat{V}_{ac,bb}\ket{\psi_{bb}^+}\right],
\end{align}
where $\text{A}$ is defined as presented in Eq.~\eqref{eq:A} of the main text. 
Substituting the previous expression into Eq.~\eqref{eq:LSbbfull2} we can then obtain 
\begin{align}
\label{eq:PsibbFull}
\ket{\psi_{bb}^+} = &\ket{\phi_{bb}^+}+\mathrm{G}^0_{bb}\left(\hat{V}_{bb,c}\mathrm{A} \left[\hat{V}_{c,bb}\ket{\psi_{bb}^+}+ \hat{V}_{c,ac}\mathrm{G}^0_{ac}\hat{V}_{ac,bb}\ket{\psi_{bb}^+}\right] + \hat{V}_{bb,ac}\mathrm{G}^0_{ac}\left[ \hat{V}_{ac,c}\mathrm{A} \left[\hat{V}_{c,bb}\ket{\psi_{bb}^+}+\right. \right. \right. \notag \\ 
&\left. \left. \left. \hat{V}_{c,ac}\mathrm{G}^0_{ac}\hat{V}_{ac,bb}\ket{\psi_{bb}^+}\right]+\hat{V}_{ac,bb}\ket{\psi_{bb}}\right] \right).
\end{align}
Equation \eqref{eq:PsibbFull} inspires us to introduce the following effective potential interaction $V_{\mathrm{eff}}$,
\begin{align}
V_{\mathrm{eff}} = &V_{bb,bb}+V_{bb,c}\mathrm{A}V_{c,bb}+V_{bb,c}\mathrm{A}V_{c,ac}\mathrm{G}^0_{ac}V_{ac,bb}+V_{bb,ac}\mathrm{G}^0_{ac}V_{ac,c}\mathrm{A}V_{c,bb}+\\ \notag 
&V_{bb,ac}\mathrm{G}^0_{ac}V_{ac,c}\mathrm{A}V_{c,ac}\mathrm{G}^0_{ac}V_{ac,bb} + V_{bb,ac}\mathrm{G}^0_{ac}V_{ac,bb}
\end{align}
which is identical to Eq.~\eqref{eq:VeffFull} of the main text.
Proceeding analogous to Sec.~\ref{subsec:ResonanceFacilitatedThreeChannelModel}, we compute $\braket{\chi|T_{\mathrm{bb},\mathrm{bb}}|\chi}$ from $\braket{\chi|V_{\mathrm{eff}}|\psi_{bb}^+}$, such that 
\begin{align}
\label{eq:T3chFull}
\mathrm{T}_{bb,bb} = &\mathrm{T}^{\mathrm{unc}}_{bb,bb}+\mathrm{S}_{\mathrm{P}}\left(\frac{\mathrm{D}}{E-\epsilon_{rr}}\left[\braket{\phi_{bb}^+|V_{bb,rr}|\phi_{rr}}\braket{\phi_{rr}|V_{rr,bb}|\psi_{bb}}+\braket{\phi_{bb}^+|V_{bb,rr}|\phi_{rr}}\braket{\phi_{rr}|V_{rr,ac}\mathrm{G}^0_{ac}V_{ac,bb}|\psi_{bb}}+ \right.\right. \notag \\
& \left.\left. \braket{\phi^+_{bb}|V_{bb,ac}\mathrm{G}^0_{ac}V_{ac,rr}|\phi_{rr}}\braket{\phi_{rr}|V_{rr,bb}|\psi_{bb}}+\braket{\phi_{bb}^+|V_{bb,ac}\mathrm{G}^0_{ac}V_{ac,rr}|\phi_{rr}}\braket{\phi_{rr}|V_{rr,ac}\mathrm{G}^0_{ac}V_{ac,bb}|\psi_{bb}}\right]\right.+ \notag \\
&\left.\braket{\phi_{bb}^+|V_{bb,ac}\mathrm{G}^0_{ac}V_{ac,bb}|\psi_{bb}}\right)
\end{align}
The previous expression can be further analysed through the application of the Gamow expansions as presented in Eqs. \eqref{eq:BBpropagator} and \eqref{eq:ACpropagator}. Here we should consider that $\braket{\Omega_{ac}|V_{ac,bb}|\psi_{bb}}$ and $\braket{\Omega_{ac}^D|V_{ac,bb}|\psi_{bb}}$ are not independent. In fact: 
\begin{align}
&\braket{\Omega_{ac}|V_{ac,bb}|\psi_{bb}} = k_{ac}^{3/2,*} \braket{\tilde{\Omega}_{ac}|V_{ac,bb}|\psi_{bb}} \qquad \text{and} \\
& \braket{\Omega_{ac}^D|V_{ac,bb}|\psi_{bb}} = k_{ac}^{3/2} \braket{\tilde{\Omega}_{ac}|V_{ac,bb}|\psi_{bb}}
\end{align} 
This means we can get rid of the factors $\braket{\Omega_{ac}^D|V_{ac,bb}|\psi_{bb}}$, $\braket{\Omega_{ac}|V_{ac,bb}|\psi_{bb}}$ and $\braket{\phi_{c}|V_{c,bb}|\psi_{bb}}$ in Eq.~\eqref{eq:T3chFull} by first multiplying Eq.~\eqref{eq:PsibbFull} with $\bra{\phi_{c}}V_{c,bb}$ and then with $\bra{\Omega_{ac}}V_{ac,bb}$. Starting with the multiplication with $\bra{\phi_{c}}V_{c,bb}$ we find 
\begin{align}
\braket{\phi_{c}|V_{c,bb}|\psi_{bb}^+} &= \braket{\phi_{c}|V_{c,bb}|\phi_{bb}^+} + \frac{\mathrm{D}}{E-\epsilon_{c}}\left(\braket{\phi_{c}|V_{c,bb}\mathrm{G}^0_{bb} V_{bb,c}|\phi_{c}}+\braket{\phi_{c}|V_{c,bb}\mathrm{G}^0_{bb}V_{bb,ac}\mathrm{G}^0_{ac}V_{ac,c}|\phi_{c}}\right)\braket{\phi_{c}|V_{c,bb}|\psi_{bb}}+ \notag \\ 
&\left(\frac{\mathrm{D}}{E-\epsilon_{c}}\left(\braket{\phi_{c}|V_{c,bb}\mathrm{G}^0_{bb}V_{bb,c}|\phi_{c}}\kappa_{ac}\braket{\phi_{c}|V_{c,ac}|\Omega_{ac}}+\braket{\phi_{c}|V_{c,bb} \mathrm{G}^0_{bb}V_{bb,ac}\mathrm{G}^0_{ac}V_{ac,c}|\phi_{c}}\kappa_{ac}\braket{\phi_{c}|V_{c,ac}|\Omega_{ac}}\right) + \right. \notag \\ 
&\left. \kappa_{ac}\braket{\phi_{cc}|V_{c,bb}\mathrm{G}^0_{bb}V_{bb,ac}|\Omega_{ac}}\right)\braket{\Omega_{ac}^D|V_{ac,bb}|\psi_{bb}} + 
\left(\frac{\mathrm{D}}{E-\epsilon_{c}}\left(-\braket{\phi_{c}|V_{c,bb}\mathrm{G}^0_{bb}V_{bb,c}|\phi_{c}}\kappa_{ac}^{\bullet}\braket{\phi_{c}|V_{c,ac}|\Omega_{ac}^D}-\right. \right. \notag \\ 
& \left. \braket{\phi_{c}|V_{c,bb} \mathrm{G}^0_{bb}V_{bb,ac}\mathrm{G}^0_{ac}V_{ac,c}|\phi_{c}}\kappa^*_{ac}\braket{\phi_{c}|V_{c,ac}|\Omega^D_{ac}}\right) -\left. \kappa_{ac}^{\bullet}\braket{\phi_{c}|V_{c,bb}\mathrm{G}^0_{bb}V_{bb,ac}|\Omega^D_{ac}}\right)\braket{\Omega_{ac}|V_{ac,bb}|\psi_{bb}},
\end{align}
where we have introduced $\kappa_{ac} = \left(2k_{ac}(\sqrt{k^2-k_{\mathrm{th}}}-k_{ac})\right)^{-1}$ and $\kappa_{ac}^{\bullet} = \left(2k_{ac}^*(\sqrt{k^2-k_{\mathrm{th}}}+k_{ac}^*)\right)^{-1}$.
Inspired by the previous equation, we define the following factors $A_1$ through $A_3$ 
\begin{align}
\label{eq:A1}
A_1 &= 1-\frac{\mathrm{D}}{E-\epsilon_{c}}\left(\braket{\phi_{c}|V_{c,bb}\mathrm{G}^0_{bb}V_{bb,c}|\phi_{c}}+\braket{\phi_{c}|V_{c,bb}\mathrm{G}^0_{bb}V_{bb,ac}\mathrm{G}^0_{ac}V_{ac,c}|\phi_{c}}\right) 
\end{align}
\begin{align}
\label{eq:A2}
A_2 = -&\left[\frac{\mathrm{D}\kappa_{ac}}{E-\epsilon_{c}}\left(\braket{\phi_{c}|V_{c,bb}\mathrm{G}^0_{bb}V_{bb,c}|\phi_{c}}\braket{\phi_{c}|V_{c,ac}|\Omega_{ac}}+ \right.\right.\notag \\
&\left.\left.\braket{\phi_{c}|V_{c,bb}\mathrm{G}^0_{bb}V_{bb,ac}\mathrm{G}^0_{ac}V_{ac,c}|\phi_{c}}\braket{\phi_{c}|V_{c,ac}|\Omega_{ac}}\right)+\kappa_{ac}\braket{\phi_{c}|V_{c,bb}\mathrm{G}^0_{bb}V_{bb,ac}|\Omega_{ac}}\right]
\end{align}
\begin{align}
\label{eq:A3}
A_3 = &\frac{\mathrm{D}\kappa^*_{ac}}{E-\epsilon_{c}}\left(\braket{\phi_{c}|V_{c,bb}\mathrm{G}^0_{bb}V_{bb,c}|\phi_{c}}\braket{\phi_{c}|V_{c,ac}|\Omega^D_{ac}}+\right.\notag \\
&\left.\braket{\phi_{c}|V_{c,bb}\mathrm{G}^0_{bb}V_{bb,ac}\mathrm{G}^0_{ac}V_{ac,c}|\phi_{c}}\braket{\phi_{c}|V_{c,ac}|\Omega_{ac}^D}\right)+\kappa_{ac}^{\bullet} \braket{\phi_{c}|V_{c,bb}\mathrm{G}^0_{bb}V_{bb,ac}|\Omega_{ac}^D},
\end{align}
such that 
\begin{align}
\label{eq:A123equation}
A_1 \braket{\phi_{c}|V_{c,bb}|\psi_{bb}^+} + A_2 \braket{\Omega_{ac}^D|V_{ac,bb}|\psi_{bb}}+A_3 \braket{\Omega_{ac}|V_{ac,bb}|\psi_{bb}} = \braket{\phi_{c}|V_{c,bb}|\phi_{bb}^+}
\end{align}
Similarly, by multiplying Eq.~\eqref{eq:PsibbFull} with $\bra{\Omega_{ac}}V_{ac,bb}$ we obtain 
\begin{align}
\braket{\Omega_{ac}|V_{ac,bb}|\psi_{bb}^+} =&  \braket{\Omega_{ac}|V_{ac,bb}|\phi_{bb}^+} +\frac{\mathrm{D}}{E-\epsilon_{c}}\left(\braket{\Omega_{ac}|V_{ac,bb}\mathrm{G}^0_{bb} V_{bb,c}|\phi_{c}}+\braket{\Omega_{ac}|V_{ac,bb}\mathrm{G}^0_{bb}V_{bb,ac}\mathrm{G}^0_{ac}V_{ac,c}|\phi_{c}}\right)\notag \\ 
&\braket{\phi_{c}|V_{c,bb}|\psi_{bb}}+\left(\frac{\mathrm{D}}{E-\epsilon_{c}}\left(\braket{\Omega_{ac}|V_{ac,bb}\mathrm{G}^0_{bb}V_{bb,c}|\phi_{c}}\kappa_{ac}\braket{\phi_{c}|V_{c,ac}|\Omega_{ac}} \right. \right. \notag \\ 
& \left. \left. + \braket{\Omega_{ac}|V_{ac,bb} \mathrm{G}^0_{bb}V_{bb,ac}\mathrm{G}^0_{ac}V_{ac,c}|\phi_{c}}\kappa_{ac}\braket{\phi_{c}|V_{c,ac}|\Omega_{ac}}\right) +  \kappa_{ac}\braket{\Omega_{ac}|V_{ac,bb}\mathrm{G}^0_{bb}V_{bb,ac}|\Omega_{ac}}\right)\notag \\ 
& \braket{\Omega_{ac}^D|V_{ac,bb}|\psi_{bb}} + \left(\frac{\mathrm{D}}{E-\epsilon_{c}}\left(-\braket{\Omega_{ac}|V_{ac,bb}\mathrm{G}^0_{bb}V_{bb,c}|\phi_{c}}\kappa_{ac}^{\bullet}\braket{\phi_{c}|V_{c,ac}|\Omega_{ac}^D}-\right. \right. \notag \\ 
& \left. \braket{\Omega_{ac}|V_{ac,bb} \mathrm{G}^0_{bb}V_{bb,ac}\mathrm{G}^0_{ac}V_{ac,c}|\phi_{c}}\kappa^*_{ac}\braket{\phi_{c}|V_{c,ac}|\Omega^D_{ac}}\right) - \notag \\ 
&\left. \kappa_{ac}^{\bullet}\braket{\Omega_{ac}|V_{ac,bb}\mathrm{G}^0_{bb}V_{bb,ac}|\Omega^D_{ac}}\right)\braket{\Omega_{ac}|V_{ac,bb}|\psi_{bb}}.
\end{align}
The previous equation expression inspires us to define the following set of parameters $C_1$ through $C_3$
\begin{align}
\label{eq:C1}
C_1 = &-\frac{\mathrm{D}}{E-\epsilon_{c}}\left(\braket{\Omega_{ac}|V_{ac,bb}\mathrm{G}^0_{bb}V_{bb,c}|\phi_{c}}+\braket{\Omega_{ac}|V_{ac,bb}\mathrm{G}^0_{bb}V_{bb,ac}\mathrm{G}^0_{ac}V_{ac,c}|\phi_{c}}\right) 
\end{align}
\begin{align}
\label{eq:C2}
C_2 = &-\left[\frac{\mathrm{D}\kappa_{ac}}{E-\epsilon_{c}}\left(\braket{\Omega_{ac}|V_{ac,bb}\mathrm{G}^0_{bb}V_{bb,c}|\phi_{c}}\braket{\phi_{c}|V_{c,ac}|\Omega_{ac}}+\right. \right. \notag \\
&\left. \left. \braket{\Omega_{ac}|V_{ac,bb}\mathrm{G}^0_{bb}V_{bb,ac}\mathrm{G}^0_{ac}V_{ac,c}|\phi_{c}}\braket{\phi_{c}|V_{c,ac}|\Omega_{ac}}\right)+\kappa_{ac}\braket{\Omega_{ac}|V_{ac,bb}\mathrm{G}^0_{bb}V_{bb,ac}|\Omega_{ac}}\right] 
\end{align}
\begin{align}
\label{eq:C3}
C_3 = &1+\frac{\mathrm{D}\kappa_{ac}^{\bullet}}{E-\epsilon_{c}}\left(\braket{\Omega_{ac}|V_{ac,bb}\mathrm{G}^0_{bb}V_{bb,c}|\phi_{c}}\braket{\phi_{c}|V_{c,ac}|\Omega_{ac}^D}+\right. \notag \\
&\left.  \braket{\Omega_{ac}|V_{ac,bb}\mathrm{G}^0_{bb}V_{bb,ac}\mathrm{G}^0_{ac}V_{ac,c}|\phi_{c}}\braket{\phi_{c}|V_{c,ac}|\Omega_{ac}^D}\right)+\kappa_{ac}^{\bullet}\braket{\Omega_{ac}|V_{ac,bb}\mathrm{G}^0_{bb}V_{bb,ac}|\Omega_{ac}^D},
\end{align}
such that
\begin{align}
\label{eq:C123equation}
C_1 \braket{\phi_{c}|V_{c,bb}|\psi_{bb}^+} + C_2 \braket{\Omega_{ac}^D|V_{ac,bb}|\psi_{bb}} + C_3 \braket{\Omega_{ac}|V_{ac,bb}|\psi_{bb}} = \braket{\Omega_{ac}|V_{ac,bb}|\phi_{bb}^+}
\end{align}
Applying Eqs. \eqref{eq:A123equation} and \eqref{eq:C123equation} we can find then find 
\begin{align}
\braket{\phi_{c}|V_{c,bb}|\psi_{bb}} &= \frac{\left(A_2 k_{ac}^{3/2}+A_3 k_{ac}^{3/2,*}\right)\braket{\Omega_{ac}|V_{ac,bb}|\phi_{bb}}-(C_2 k_{ac}^{3/2}+C_3 k_{ac}^{3/2,*})\braket{\phi_{c}|V_{c,bb}|\phi_{bb}}}{(A_2 C_1 - A_1 C_2)k_{ac}^{3/2}+(A_3 C_1-A_1 C_3) k_{ac}^{3/2,*}} \\[15pt]
\braket{\tilde{\Omega}_{ac}|V_{c,bb}|\psi_{bb}} &= \frac{C_1 \braket{\phi_{c}|V_{c,bb}|\phi_{bb}}-A_1 \braket{\Omega_{ac}|V_{ac,bb}|\phi_{bb}}}{(A_2 C_1 - A_1 C_2)k_{ac}^{3/2}+(A_3 C_1-A_1 C_3) k_{ac}^{3/2,*}},
\end{align}
with $\braket{\Omega_{ac}|V_{c,bb}|\psi_{bb}} = k_{ac}^{3/2,*} \braket{\tilde{\Omega}_{ac}|V_{c,bb}|\psi_{bb}}$ and $\braket{\Omega^D_{ac}|V_{c,bb}|\psi_{bb}} = k_{ac}^{3/2} \braket{\tilde{\Omega}_{ac}|V_{c,bb}|\psi_{bb}} $. 
The previous two expressions can be directly substituted into Eq.~\eqref{eq:T3chFull} in order to obtain the full three-channel T-matrix in terms of the direct scattering wave functions $\phi^+_{bb}$ and the bound state wave function $\phi_c$.
\end{widetext}
\end{document}